\documentclass[twocolumn,amsmath,amssymb,prb]{revtex4}
\usepackage{color,epsfig,rotating,graphicx,subfigure}

\usepackage[latin1]{inputenc}
\usepackage[english]{babel}
\usepackage{dsfont}
\usepackage{textcomp}

\newcommand{\ri}{\text{i}}
\newcommand{\re}{{\rm e}}
\newcommand{\rd}{{\rm d}}
\newcommand{\rr}{{\rm r}}

\newcommand{\dr}{\!\! \rm{d}^3 r}

\renewcommand{\Im}{\text{Im}}
\renewcommand{\Re}{\text{Re}}

\renewcommand{\ll}{\big \langle \!\! \big \langle}
\renewcommand{\rr}{\big \rangle \!\! \big \rangle}

\graphicspath{{./Figures/}}

\begin{document}

\title{Dipole model for far-field thermal emission of a nanoparticle above a planar substrate}

\date{\today}

\author{Florian Herz}
\affiliation{Institut f{\"u}r Physik, Carl von Ossietzky Universit{\"a}t, D-26111 Oldenburg, Germany}
\author{Svend-Age Biehs}
\email{s.age.biehs@uni-oldenburg.de}
\affiliation{Institut f{\"u}r Physik, Carl von Ossietzky Universit{\"a}t, D-26111 Oldenburg, Germany}

\begin{abstract}
	We develop a dipole model describing the thermal far-field radiation of a nanoparticle in close vicinity to a substrate. By including in our description the contribution of eddy currents and the possibility to choose different temperatures for the nanoparticle, the substrate, and the background, we generalize the existing models. We discuss the impact of the different temperatures, particle size, emission angle, and the distance dependence for all four combinations of gold and SiC nanoparticles or substrates.
\end{abstract}

\maketitle

\section{Introduction}

The theoretical problem of thermal emission of a nanoparticle in the vicinity of a planar substrate within the dipole model together with the framework of fluctuational electrodynamics has a relatively long history. In the first works it was used to determine the exchange of thermal radiation between a spherical nanoparticle and a planar substrate when bringing the nanoparticle into the near-field of the substrate~\cite{Dorofeyev97,Dorofeyev98,JPendry99,MuletEtAl01}. In subsequent works, the contribution due to multiple interactions between the nanoparticle and the surface~\cite{VolokitinPersson07}, the contribution of eddy currents~\cite{DedkovKyasov07,ChapuisEtAl08b}, and mixed magnetic and dielectric contributions~\cite{Joulain2} where added to the description, as well as the generalization to ellipsoidal nanoparticles~\cite{HuthEtAl2010}, the inclusion of surface roughness for the substrate~\cite{Biehs2010}, and multipolar moments for the nanoparticle~\cite{Doro2008}. In a similar manner also the scattering of the thermal near-field by a nanoparticle into the far-field~\cite{Joulain2,Jarzembski,Herz} and the direct thermal emission of a nanoparticle into the far-field~\cite{Joulain2,Asheichyk2017,Herz} has been studied.

This interest into thermal emission is triggered by several infrared near-field thermal imaging setups where typically a sharp tip of an atomic force microscope (AFM) cantilever, scattering optical microscope (SNOM), or scanning tunneling microscope (STM) is brought close to a surface~\cite{DeWilde,Kittel2008,WorbesEtAl2013,Huth2011,Jones,WengEtAl2018}. For example, in the so-called thermal radiation scanning tunneling microscope (TRSTM)~\cite{DeWilde}, thermal radiation of a heated surface is scattered by a sharp tip into the far-field where its spectrum can be measured~\cite{Babuty}. A similar experiment can be made with the thermal infrared near-field spectroscope (TINS) \cite{Jones, O'Callahan} which is based on an AFM probe and allows for heating of the surface and the tip with respect to the environment. The so-called scanning noise microscope (SNoiM) is a highly improved TRSTM which does not need to heat the tip or the substrate to get a signal like for TRSTM or TINS \cite{Lin,WengEtAl2018, Komiyama} due to an ultra-sensitive single-photon detector working at $4.2$ K. On the other hand, the SNoiM is measuring the heat flux in a narrow wavelength band around $14.2\,\mu{\rm m}$ and in its present state it cannot measure broad spectra like the TINS or TRSTM. A simple approach to model such experimental setups is to approximate the foremost part of the probe as a nanoparticle~\cite{Joulain2, Jarzembski,Herz}. Then, in lowest order one can attribute the main influence on the signal due to the local density of states (LDOS) at the tip position~\cite{DeWilde, Jones, O'Callahan}. Although a comparison to the LDOS can explain many properties of the measured spectra, this model, for example, cannot explain the red-shift of the surface phonon polariton (SPhP) frequency for increasing distances between a Si tip and a SiC substrate \cite{Jones}. Furthermore, the modeling of these experiments by the scattering of the near-field by a nanoparticle neglects at least one of the following two facts: First, apart from the induced electric dipole moment, eddy currents can emerge, too. Those eddy currents are particularly important for metallic particles \cite{DedkovKyasov07,ChapuisEtAl08b, Dong} and were included in the dipole model in Ref.~\cite{Joulain2}. Second, in the experiments the tip of the probes, the substrate, and, additionally, the background can have different temperatures. Here, we provide a generalization of the model~\cite{Joulain2,Herz} to take these different temperatures into account. Furthermore, instead of neglecting divergent terms of the Green function appearing in the dipole model, we conduct a renormalization procedure as described for the coupled dipole model in Ref.~\cite{Lakhtakia}.

Our work is organized as follows. In Sec.~II we introduce the renormalization method and the dressed polarizabilities. In Sec.~III we determine the analytical expressions for the power emitted by the nanoparticle, the surface and the mixed terms. In Sec.~IV we discuss the numerical example of a Au or SiC nanoparticle above a Au or SiC substrate. Finally, in Sec.~V we summarize our results and give a conclusion.

\section{Fluctuational fields and dressed polarizability}

\begin{figure}[hbt]
    \center
    \includegraphics[width=0.45\textwidth]{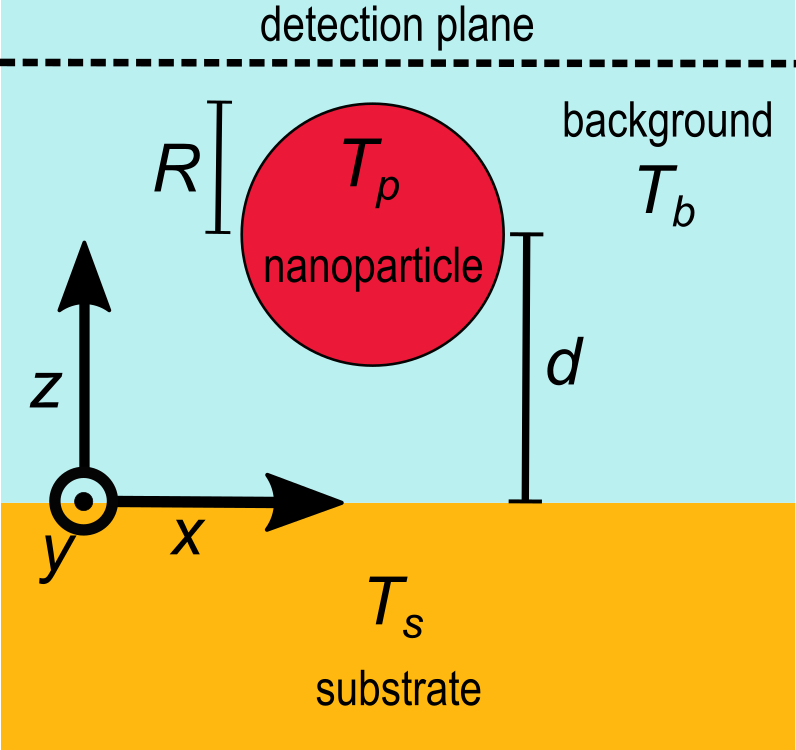}
    \caption{Sketch of the considered configuration of a nanoparticle with radius $R$ at temperature $T_\text{p}$ and distance $d$ above a planar substrate at temperature $T_\text{s}$ immersed in a background with temperature $T_\text{b}$. The whole setup is considered to be underneath a detection plane through which the heat flux is measured.}
    \label{fig:Setup}
\end{figure}

Consider the following setup as sketched in Fig.~\ref{fig:Setup}: A nanoparticle with radius $R$ and temperature $T_\text{p}$ is placed at distance $d$ to a semi-infinite planar substrate at $z \leq 0$ having a temperature $T_\text{s}$. The materials of the nanoparticle and substrate are assumed to be non-magnetic, homogeneous, and isotropic. They are embedded in vacuum which is filled with a thermal fluctuating electromagnetic field at temperature $T_\text{b}$ due to the thermal photons in the vacuum part. Hence, we assume that the substrate, the nanoparticle, and the backround can be considered to be in local thermal equilibrium at the given temperatures. Furthermore, we will assume that the  fields due to the thermal sources in the substrate and nanoparticle and the background fields are statistically independent. We want to determine the spectral power of heat radiation emitted into the far-field through a plane parallel to the substrate at distance $z > d + R$, i.e.\ above the nanoparticle (see detection plane in Fig.~\ref{fig:Setup}). Due to the translational symmetry in x-y direction or rotational symmetry with respect to the z-axis, only the z-component of the mean Poynting vector of the total thermal fluctuational fields will be of interest. 

To determine the mean Poynting vector in this configuration, the expressions of the total electric and magnetic fields are required. It is possible to obtain these expressions by solving the volume-integral equation~\cite{Hecht} 
\begin{align}
   \underline{\mathbf{F}}_\text{tot} (\mathbf{r}) & = \underline{\mathbf{F}}_\text{env} (\mathbf{r}) + \ri \mu_0 \omega \int_{V_\text{p}} \dr' \underline{\underline{\mathds{G}}} (\mathbf{r}, \mathbf{r}') \underline{\mathbf{J}}  (\mathbf{r}') 
\label{eq:lse}
\end{align}
introducing angular frequency $\omega$, vacuum permeability $\mu_0$, and imaginary unit $\ri$. The total field $\underline{\mathbf{F}}_\text{tot}$  consists of the field of the environment $\underline{\mathbf{F}}_\text{env}$ and the field generated by the current densities $\underline{\mathbf{J}}  (\mathbf{r}')$ inside the nanoparticle. Hence, the volume integral is carried out over the volume $V_\text{p}$ of the nanoparticle. Here, we take the electric and magnetic source currents into account in order to describe dielectric and magnetic nanoparticles~\cite{DedkovKyasov07,ChapuisEtAl08b, HuthEtAl2010} so that the above quantities are given by the block vectors
\begin{align}
  \underline{\mathbf{F}} (\mathbf{r}) & = \left( \mathbf{E} (\mathbf{r}) , \mathbf{H} (\mathbf{r}) \right)^t, 
  \label{eq:F} \\
  \underline{\mathbf{J}} (\mathbf{r}) & = \left( \mathbf{J}_\text{E} (\mathbf{r}) , \mathbf{J}_\text{H} (\mathbf{r}) \right)^t, 
  \label{eq:J} 
\end{align}
and block matrix
\begin{equation}
  \underline{\underline{\mathds{G}}} (\mathbf{r}, \mathbf{r}') = \begin{pmatrix} \mathds{G}_\text{EE} (\mathbf{r}, \mathbf{r}') & \mathds{G}_\text{EH} (\mathbf{r}, \mathbf{r}') \\ \mathds{G}_\text{HE} (\mathbf{r}, \mathbf{r}') & \mathds{G}_\text{HH} (\mathbf{r}, \mathbf{r}') \end{pmatrix}.
\end{equation}
Note that the environment fields with index ``env'' include both the fields of the background and the substrate. 

The volume-integral equation is typically simplified by using the dipole approximation which is implemented by replacing the current densities $\underline{\mathbf{J}}$ by dipole moments $\underline{\mathbf{d}}$ located at the position $\mathbf{r}_\text{p}$ of the nanoparticle's center via the relation $\underline{\mathbf{J}}(\mathbf{r}) = - \ri \omega \underline{\mathbf{d}} \delta(\mathbf{r} -\mathbf{r}_\text{p})$. The disadvantage of this implementation is that it leads to the necessity for evaluating the Green's functions $\underline{\underline{\mathds{G}}}(\mathbf{r}_\text{p}, \mathbf{r}_\text{p})$ which can be decomposed into a vacuum part and a scattering part $\underline{\underline{\mathds{G}}} = \underline{\underline{\mathds{G}}}_{\rm vac} + \underline{\underline{\mathds{G}}}_{\rm sc}$. It is well known that the real part of the vacuum Green's function diverges in such cases~\cite{Yaghjian} and is, therefore, very often neglected. One simply reimplements this part by renormalizing the polarizability of the nanoparticle leading to a ``dressed'' polarizability as detailed in Ref.~\cite{Messina}. Here, we will follow another route of employing the dipole or long wavelength approximation (LWA) as discussed in detail in Refs.~\cite{Lakhtakia,Albaladejo}. To this end, we first derive the fields $\underline{\mathbf{F}}_{\text{in}}$ inside the nanoparticle by consider the nanoparticle as a finite sized subwavelength sphere, i.e.\ for propagating waves the sphere must be much smaller than the wavelength and much smaller than the decay length of the field for evanescent modes. In this case, it is reasonable to assume that the current density is approximately constant within this very small finite volume~\cite{Lakhtakia,Albaladejo} and can, therefore, be expressed as
\begin{equation}
  \underline{\mathbf{J}} (\mathbf{r}) = - \ri \frac{\omega}{V_p} \underline{\mathbf{d}}
\end{equation}
where $\underline{\mathbf{d}} = (\mathbf{p} , \mathbf{m})^t$ contains the electric and magnetic dipole moments $\mathbf{p}$ and $\mathbf{m}$ corresponding to the entries $\mathbf{J}_\text{E}$ and $\mathbf{J}_{\text{H}}$. Therefore, for $\mathbf{r} \in V_\text{p}$ we can take the current density in the second term in Eq.~(\ref{eq:lse}) out of the volume integral so that only the Green's function is integrated. This integral can be understood as a volume average of the function defined by
\begin{align}
	\langle \underline{\underline{\mathds{G}}} (\mathbf{r}) \rangle & = \frac{1}{V_{\rm p}} \int_{V_{\rm p}} \dr' \underline{\underline{\mathds{G}}} (\mathbf{r}, \mathbf{r}') .
\end{align}
Hence, the volume-integral equation for $\mathbf{r} \in V_\text{p}$ is
\begin{equation}
	\underline{\mathbf{F}}_\text{in} (\mathbf{r}) = \underline{\mathbf{F}}_\text{env} (\mathbf{r}) + \mu_0 \omega^2 \langle \underline{\underline{\mathds{G}}} (\mathbf{r}) \rangle \underline{\mathbf{d}}. 
\label{eq:lsein}
\end{equation}
Now, the dipole moment can be separated into a fluctuating and induced contribution due to the fields inside the nanoparticle volume by
\begin{equation}
	\underline{\mathbf{d}} = \underline{\mathbf{d}}_\text{fl} + \varepsilon_0 V \underline{\underline{\tilde{\boldsymbol{\chi}}}} \underline{\mathbf{F}}_\text{in} (\mathbf{r})
  \label{eq:pm} 
\end{equation}
with the vacuum permittivity $\varepsilon_0$ and 
\begin{equation}
  \underline{\underline{\tilde{\boldsymbol{\chi}}}} = \begin{pmatrix} \chi_\text{E} & 0 \\ 0 & \chi_\text{H} \end{pmatrix}. 
\end{equation}
Here, $\chi_\text{E/H}$ are the electric material response to the electric/magnetic field. In conclusion, we can express the fields inside the nanoparticle by the closed formula (omitting the arguments for convenience)
\begin{align}
  \underline{\mathbf{F}}_\text{in} & = \left[ \mathds{1} - k_0^2 V \underline{\underline{\tilde{\boldsymbol{\chi}}}} \langle \underline{\underline{\mathds{G}}} \rangle \right]^{-1} \left[ \underline{\mathbf{F}}_\text{env} + \mu_0 \omega^2 \langle \underline{\underline{\mathds{G}}} \rangle \underline{\mathbf{d}}_\text{fl} \right]
\label{eq:felder1}
\end{align}
defining the vacuum wave number $k_0 = \omega/c$. 
Inserting Eq. \eqref{eq:felder1} into Eq. \eqref{eq:pm}, we get
\begin{align}
  \underline{\mathbf{d}} & = \left[ \mathds{1} + k_0^2 \underline{\underline{\boldsymbol{\alpha}}} \langle \underline{\underline{\mathds{G}}}_\text{EE} (\mathbf{r}_\text{p}) \rangle \right] \underline{\mathbf{d}}_\text{fl} + \varepsilon_0 \underline{\underline{\boldsymbol{\alpha}}} \, \underline{\mathbf{F}}_\text{env} (\mathbf{r}_\text{p}).
\label{eq:dipolmoment2}
\end{align}
where we have introduced the polarizability block matrix
\begin{equation}
	\underline{\underline{\boldsymbol{\alpha}}} = V \underline{\underline{\tilde{\boldsymbol{\chi}}}}   \left[ \mathds{1} - k_0^2 V \underline{\underline{\tilde{\boldsymbol{\chi}}}} \langle \underline{\underline{\mathds{G}}} (\mathbf{r}_\text{p}) \rangle \right]^{-1} \equiv \begin{pmatrix} \boldsymbol{\alpha}_\text{EE} & \boldsymbol{\alpha}_\text{EH} \\ \boldsymbol{\alpha}_\text{HE} & \boldsymbol{\alpha}_\text{HH} \end{pmatrix} .
\label{eq:pol}
\end{equation}
The components of this polarizability block matrix are
\begin{align}
  \boldsymbol{\alpha}_\text{EE} & = \frac{V \chi_\text{E} \mathbf{e}_\perp \otimes \mathbf{e}_\perp}{1 - k_0^2 V \chi_\text{E} \langle G_{\text{EE},\perp} (\mathbf{r}_\text{p}) \rangle + \frac{k_0^4 V^2 \chi_\text{E} \chi_\text{H} \langle G_\text{HE} (\mathbf{r}_\text{p}) \rangle^2}{1 - k_0^2 V \chi_\text{H} \langle G_{\text{HH},\perp} (\mathbf{r}_\text{p}) \rangle}} \notag \\
  & \quad + \frac{V \chi_\text{E} \mathbf{e}_z \otimes \mathbf{e}_z}{1 - k_0^2 V \chi_\text{E} \langle G_\text{EE,z} (\mathbf{r}_\text{p}) \rangle} , 
  \label{eq:alpha_EE} \\
  \boldsymbol{\alpha}_\text{EH} & = \frac{k_0^2 V \chi_\text{H/E} \alpha_{\text{EE/HH},\perp} \langle G_\text{HE} (\mathbf{r}_\text{p}) \rangle}{1 - k_0^2 V \chi_\text{H/E} \langle G_{\text{HH/EE},\perp} (\mathbf{r}_\text{p}) \rangle} 
	\mathds{X} 
\end{align}
with $\mathbf{e}_\perp \otimes \mathbf{e}_\perp = \mathbf{e}_x \otimes \mathbf{e}_x + \mathbf{e}_y \otimes \mathbf{e}_y$
and $\mathds{X} = \mathbf{e}_x \otimes \mathbf{e}_y - \mathbf{e}_y \otimes \mathbf{e}_x$.

Furthermore, $\boldsymbol{\alpha}_\text{HE}  = \boldsymbol{\alpha}_\text{EH}$ holds and $\boldsymbol{\alpha}_\text{HH}$ follows from $\boldsymbol{\alpha}_\text{EE}$ by interchanging the field indices. Note that $\boldsymbol{\alpha}_{\rm EE}^t =\boldsymbol{\alpha}_{\rm EE}$, $\boldsymbol{\alpha}_{\rm HH}^t =\boldsymbol{\alpha}_{\rm HH}$, and $\boldsymbol{\alpha}_{\rm EH}^t = - \boldsymbol{\alpha}_{\rm EH} $. The explicit expressions for the volume averaged Green's functions can be found in appendix~\ref{App:VolumeAverage}. These four polarizabilities are the general ``dressed polarizabilities'' because they are not the polarizabilities of the isolated nanoparticle but also contain information about the interaction with the substrate material by means of the scattering part of the Green's function. In its structure the purely electric part resembles the results in recent works~\cite{Messina, Albaladejo}. By the magnetic contribution we retrieved the results for recently derived dressed polarizabilities~\cite{Joulain2} apart from the missing vacuum contribution which is included in our approach. 

With the expression of the dipole moment in terms of the dressed polarizability we can now derive the analytical expression for the field outside the nanoparticle. For $\mathbf{r} \notin V_\text{p}$ the volume average of the Green's function is simply $\langle \underline{\underline{\mathds{G}}} (\mathbf{r}) \rangle = \underline{\underline{\mathds{G}}}(\mathbf{r},\mathbf{r}_\text{p})$~\cite{Albaladejo} within the long-wavelength approximation. Hence, the field outside the nanoparticle  Eq.~(\ref{eq:lse}) can be written as
\begin{equation}
   \underline{\mathbf{F}}_\text{out} (\mathbf{r}) = \underline{\mathbf{F}}_\text{env} (\mathbf{r}) + \mu_0 \omega^2 \underline{\underline{\mathds{G}}} (\mathbf{r}, \mathbf{r}_\text{p}) \underline{\mathbf{d}}.
\end{equation}
Inserting Eq.~\eqref{eq:dipolmoment2}, we obtain
\begin{equation}
  \begin{split}
    \underline{\mathbf{F}}_\text{out} (\mathbf{r}) & = \mu_0 \omega^2 \underline{\underline{\mathds{G}}} (\mathbf{r}, \mathbf{r}_\text{p}) \left[ \mathds{1} + k_0^2 \underline{\underline{\boldsymbol{\alpha}}} \langle \underline{\underline{\mathds{G}}} (\mathbf{r}_\text{p}) \rangle \right] \underline{\mathbf{d}}_\text{fl} \\
    & \quad + k_0^2 \underline{\underline{\mathds{G}}} (\mathbf{r}, \mathbf{r}_\text{p}) \underline{\underline{\boldsymbol{\alpha}}} \, \underline{\mathbf{F}}_\text{env} (\mathbf{r}_\text{p}) + \underline{\mathbf{F}}_\text{env} (\mathbf{r}).
  \end{split}
\label{Eq:Fout}
\end{equation}
This is the final result for the fields outside the nanoparticle expressing it in terms of the fluctuational environmental field and the fluctuating dipole moments. The first term describes the directly emitted field of the nanoparticle. The scattering of the environmental field at the nanoparticle is included in the second contribution. The last term solely corresponds to the environmental field $ \underline{\mathbf{F}}_\text{env} (\mathbf{r}) = \underline{\mathbf{F}}_\text{s} (\mathbf{r}) + \underline{\mathbf{F}}_\text{b} (\mathbf{r})$ incorporating the substrate and background field evaluated at position $\mathbf{r}$. 

\section{Power emitted into the far field}

The power emitted into the far-field in the configuration in Fig.~\ref{fig:Setup} is given by the integration of the z-component of the Poynting vector over a x-y plane above the nanoparticle. Formally, the spectral power is given by (Einstein notation)
\begin{equation}
\begin{split}
	P_\omega^{\rm tot} &= \int \rd^2 x\,  \ll  \mathbf{S}^\text{out} \rr^\omega \cdot \mathbf{e}_z  \\
                           &= 2 \Re \int \rd^2 x\, \epsilon_\emph{ijz} \ll \mathbf{E}_\text{out} (\mathbf{r}) \otimes \mathbf{H}_\text{out}^\dagger (\mathbf{r}) \rr^\omega_{ij} 
\end{split}
\label{eq:P_gen}
\end{equation}
where $\epsilon_{ijk}$ denotes the Levi-Civita symbol, $\,\, \rd^2 x$ is an infinitesimal area element of the x-y detection plane which is indicated in Fig.~\ref{fig:Setup}, and $\ll \cdot \rr^\omega$ defines the stochastic average in frequency space. By inserting the fields from Eq.~(\ref{Eq:Fout}) and assuming that the fluctuational fields of the substrate, background and nanoparticle are statistically independent, we obtain
\begin{equation}
	P_\omega^{\rm tot} = P_\omega^s + P_\omega^{\rm env} + P_\omega^{\rm np}.
\end{equation}
Here, 
\begin{equation}
   P_\omega^s = 2 \Re \int \rd^2 x\, \epsilon_\emph{ijz} \ll \mathbf{E}_\text{env} (\mathbf{r}) \otimes \mathbf{H}_\text{env}^\dagger (\mathbf{r}) \rr^\omega_{ij}
\end{equation}
is the thermal radiation of the substrate at temperature $T_\text{s}$ into its environment at temperature $T_\text{b}$ without the nanoparticle and its analytical expression is well known~\cite{Polder}. On the other hand $P_\omega^{\rm np}$, including all the terms with correlation functions of the fluctuational dipole moments $\underline{\mathbf{d}}_{\rm fl}$, describes the thermal emission of the nanoparticle in the presence of the substrate. Finally, $P_\omega^{\rm env}$ contains all the other terms including correlation functions of the environment fields and incorporates, therefore, corrections to the thermal emission. For example, it includes the power emitted by the substrate which is absorbed by the nanoparticle. Note that the thermal emission of the nanoparticle into the substrate, as measured in near-field imaging experiments in Refs.~\cite{Kittel2008,WorbesEtAl2013}, can be evaluated in a similar manner by integrating the mean Poynting vector over a plane between the particle and the substrate. However, here we focus on the far-field emission, only.

In the following we will evaluate all contributions to the total power. To this end, it is necessary to deal with the field correlation functions of the environment fields $\mathds{C}_\text{env}^\text{FF'} = \ll \mathbf{F}\otimes\mathbf{F}'\rr$ where F and F' symbolize the electric or magnetic fields. Due to the statistical independence of the substrate and background fields, we can separate the correlation function into a substrate and background part
\begin{align}
  \mathds{C}_\text{env}^\text{FF'} (\mathbf{r}, \mathbf{r}', \omega) & = \mathds{C}_\text{s}^\text{FF'} (\mathbf{r}, \mathbf{r}', \omega) + \mathds{C}_\text{b}^\text{FF'} (\mathbf{r}, \mathbf{r}', \omega) .
\label{eq:C_sep}
\end{align}
The fluctuational fields of the substrate are generated by thermal sources in the substrate volume $V_\text{s}$. Hence, we can assume local equilibrium for the substrate at temperature $T_\text{s}$ and evaluate these correlation function using Rytov's formulae giving\cite{Eckhardt}
\begin{align}
  \mathds{C}^\text{FF'}_\text{s,leq} (\mathbf{r}, \mathbf{r}') & = 2 \Theta_\text{s} k_0^2 \mu_0 \omega \Im(\varepsilon_\text{s}) \notag \\
                                                               & \quad \times \int_{V_\text{s}} \dr'' \mathds{G}_\text{FE}^\text{s} (\mathbf{r}, \mathbf{r}'')   \mathds{G}_\text{F'E}^{\text{s} \dagger} (\mathbf{r}', \mathbf{r}'') 
\label{eq:leqs} .
\end{align}
We employ the definition $\Theta_\gamma = \hbar \omega \bigl( n_\gamma + \frac{1}{2}\bigr)$ where $n_\gamma$ is the Bose-Einstein occupation number ($\gamma = {\rm s, p, b}$) 
\begin{align}
  n_\gamma & = \left(e^{\frac{\hbar \omega}{k_B T_\gamma}} - 1 \right)^{-1}
\end{align}
with the Boltzmann constant $k_{\rm B}$ and the reduced Planck constant $\hbar$. The Green's functions used here are those with source points within the substrate and observation points outside the substrate as explicitely given in appendix~\ref{App:GreenDyads}. 

To evaluate the correlation functions for the background fields, we recall the expressions for global thermal equilibrium~\cite{Agarwal,Eckhardt}
\begin{align}
  \mathds{C}^\text{EE/HH}_\text{env,eq} (\mathbf{r}, \mathbf{r}') & = 2 \Theta \mu_0 \omega \Im \left[\mathds{G}_\text{EE/HH} (\mathbf{r}, \mathbf{r}')\right] , \label{eq:env1} \\
    \mathds{C}^\text{EH}_\text{env,eq} (\mathbf{r}, \mathbf{r}') & = - 2 \ri \Theta \mu_0 \omega \Re \left[\mathds{G}_\text{EH} (\mathbf{r}, \mathbf{r}')\right] .
\label{eq:env3}
\end{align}
To obtain an expression for the background contributions in local thermal equilibrium, we exploit the fact that, when $T_\text{s} = T_\text{p}  = T_\text{b}$,  Eq.~\eqref{eq:C_sep} must fulfill these expressions for global equilibrium. Then, the desired background correlation can be evaluated from Eqs. \eqref{eq:leqs}-\eqref{eq:env3}. Thus, we obtain, for example, for the correlation function with two electric fields
\begin{align}
  \mathds{C}^\text{EE}_\text{b,leq} (\mathbf{r}, \mathbf{r}') & = 2 \Theta_\text{b} \mu_0 \omega \Bigl[ \Im \left[\mathds{G}_\text{EE} (\mathbf{r}, \mathbf{r}')\right] \notag \\
   & \quad - k_0^2 \Im(\varepsilon_\text{s}) \int_{V_\text{s}} \dr'' \mathds{G}_\text{EE}^\text{s} (\mathbf{r}, \mathbf{r}'')   \mathds{G}_\text{EE}^{\text{s} \dagger} (\mathbf{r}', \mathbf{r}'')  \Bigr] .
\end{align}
Combining both the substrate and background correlation functions, we exemplary end up with 
\begin{equation}
  \begin{split}
    \mathds{C}^\text{EE}_\text{env} (\mathbf{r}, \mathbf{r}') & = 2 \Theta_\text{b} \mu_0 \omega \Im \left[\mathds{G}_\text{EE} (\mathbf{r}, \mathbf{r}')\right] \\
    & \quad + 2 \left( \Theta_\text{s} - \Theta_\text{b} \right) k_0^2 \mu_0 \omega  \Im(\varepsilon_\text{s})  \\
    & \quad \times \int_{V_\text{s}} \dr'' \mathds{G}_\text{EE}^\text{s} (\mathbf{r}, \mathbf{r}'')   \mathds{G}_\text{EE}^{\text{s} \dagger} (\mathbf{r}', \mathbf{r}'') .
  \end{split}
\end{equation}
Similar expressions hold for the other correlation functions. Hence, the correlation functions $\mathds{C}^\text{FF'}_\text{env}$ always consist of a global equilibrium term and a local equilibrium correction vanishing for $T_\text{s} = T_\text{b}$. Therefore, we can also separate the emitted power into exactly these two contributions
\begin{equation}
  P_\omega^{\rm env} = P_\omega^\text{eq} + P_\omega^\text{leq}.
\end{equation}
We will refer to $P_\omega^\text{eq}$ by ``equilibrium contribution'' (EQC) and to $P_\omega^\text{leq}$ by ``local equilibrium contribution'' (LEQC). 

In the following, we provide the explicit expressions for the different terms. The special cases of these expressions for a perfect metallic or black body substrate are given in appendices \ref{App:PM} and \ref{App:BB}

\subsection{Contribution of the substrate}

First, we recall the known expression for the thermal radiation 
of the planar substrate into the background. It reads~\cite{Polder}
\begin{equation}
   P_\omega^\text{s} = A \bigl( \Theta_\text{s} - \Theta_\text{b} \bigr) \sum_{k = \text{E,H}} \int_0^{k_0}\!\! \frac{\rd k_\perp}{2 \pi} k_\perp (1 - |r_k|^2)
\label{Eq:Psubstrate}
\end{equation}
where $r_\text{E}$ and $r_\text{H}$ are the Fresnel coefficients for p- and s-polarized light as defined in appendix~\ref{App:GreenDyads} and $A$ is the area of a plane above the substrate through which the heat flux is flowing.

\subsection{EQC}

Using the expressions for the correlation functions, we find
\begin{equation}
  \begin{split}
     P_{\omega}^\text{eq} & = - 4 \Theta_\text{b}\mu_0 \omega \epsilon_{ijz} \int \rd^2 x\,  \\
& \quad \times \Re \Bigl[ \mathds{G}_{\text{E} k} (\mathbf{r}, \mathbf{r}_\text{p})   \Bigl[ \boldsymbol{\chi}^k + \mathds{F}_k   \boldsymbol{\chi}^{\bar{k}}   \mathds{F}_k^\dagger \Bigr]   \mathds{G}_{\text{H} k}^\dagger (\mathbf{r}, \mathbf{r}_\text{p}) \\
& \quad + \mathds{G}_{\text{E} k} (\mathbf{r}, \mathbf{r}_\text{p})   \Bigl[ \boldsymbol{\chi}^k   \mathds{F}_{\bar{k}}^\dagger + \mathds{F}_k   \boldsymbol{\chi}^{\bar{k}} \Bigr]   \mathds{G}_{\text{H} \bar{k}}^\dagger (\mathbf{r}, \mathbf{r}_\text{p}) \Bigr]_{ij}
  \end{split}
\end{equation}
where we introduced the generalized susceptibility block matrix (Einstein summation over $k = \text{E,H}$) 
\begin{equation}
\begin{split}
  \boldsymbol{\chi}^k & = k_0^2 \Im \left( \boldsymbol{\alpha}_{kk} \right) - k_0^4 \boldsymbol{\alpha}_{kk}   \Im \Bigl(\langle \mathds{G}_{kk} (\mathbf{r}_\text{p}) \rangle \\
  & \quad + \langle \mathds{G}_\text{HE} (\mathbf{r}_\text{p}) \rangle   \mathds{F}_{\bar{k}} \Bigr)   \boldsymbol{\alpha}_{kk}^\dagger , 
  \end{split}
  \label{eq:chi_eh}
\end{equation}
analogue to the one introduced for the purely electric case~\cite{Krueger, Herz2}, and
\begin{equation}
  \mathds{F}_{k} = k_0^2 V \chi_k \left[ \mathds{1} - k_0^2 V \chi_k \langle \mathds{G}_{kk} (\mathbf{r}_\text{p}) \rangle \right]^{-1}   \langle \mathds{G}_\text{HE} (\mathbf{r}_\text{p}) \rangle .
\end{equation}
The bar notes the ``inversion'' $\bar{\text{E}} = \text{H}$ and $\bar{\text{H}} = \text{E}$.  

Finally, by inserting the Green's functions from appendix~\ref{App:GreenDyads} and a lengthy calculation we find
\begin{equation}
\begin{split}
    P_\omega^\text{eq} & = - k_0 \Theta_\text{b} \Bigl[ \left(\chi_{\text{E},\perp} + |F_\text{E}|^2 \chi_{\text{H},\perp} \right) I_{\text{E},\perp}^\text{pr} + \chi_{\text{E},z} I_{\text{E},z}^\text{pr}  \\
      & \quad + \frac{\varepsilon_0}{\mu_0} \left[ \left(\chi_{\text{H},\perp} + |F_\text{H}|^2 \chi_{\text{E},\perp} \right) I_{\text{H},\perp}^\text{pr} + \chi_{\text{H},z} I_{\text{H},z}^\text{pr} \right]  \\
      & \quad + 2 \sqrt{\frac{\varepsilon_0}{\mu_0}} \Re \Bigl[ \left( \chi_{\text{H},\perp} F_\text{E} - \chi_{\text{E},\perp} F_\text{H}^{*} \right) I_{\text{c}}^\text{pr} \Bigr] \Bigr]
  \end{split}
\end{equation}
with 
\begin{align}
  I_{k,\perp}^\text{pr} & = \int_0^{k_0} \frac{\text{d} k_\perp}{2 \pi} \frac{k_\perp}{k_z k_0} \biggl( \frac{k_z^2}{k_0^2} \big |1 - r_k e^{2 \text{i} k_z d} \big|^2 \notag \\
     & \quad + \big|1 + r_{\bar{k}} e^{2 \text{i} k_z d} \big|^2 \biggr) 
     \label{eq:I_p_perp} , \\
  I_{k,z}^\text{pr} & = \int_0^{k_0} \frac{\text{d} k_\perp}{2 \pi} \frac{k_\perp^3}{k_0^3 k_z} \big|1 + r_k e^{2 \text{i} k_z d} \big|^2 
  \label{eq:I_p_z}
\end{align}
and 
\begin{equation}
  \begin{split}
     I_{\text{c}}^\text{pr} & = \int_0^{k_0} \frac{\text{d} k_\perp}{2 \pi} \frac{k_\perp}{k_0^2} \bigl[ \left( 1 + r_\text{H} e^{2 \ri k_z d} \right) \left(1 - r_\text{H} e^{2 \ri k_z d}  \right)^{*} \\
        & \quad + \left( 1 - r_\text{E} e^{2 \ri k_z d} \right) \left( 1 + r_\text{E} e^{2 \ri k_z d} \right)^{*} \bigr] .
  \end{split}
\end{equation}

\subsection{Nanoparticle contribution}

By employing the fluctuation-dissipation theorem for the electric and magnetic dipole moments
\begin{align}
	\ll \mathbf{p}_\text{fl} \otimes \mathbf{p}_\text{fl}^\dagger \rr^\omega & = \frac{2}{\mu_0 \omega^3} \Theta_\text{p} \mathds{A}_\text{EE}^{-1}   \boldsymbol{\chi}^\text{E}   \mathds{A}_\text{EE}^{-1 \dagger},\\ 
        \ll \mathbf{m}_\text{fl} \otimes \mathbf{m}_\text{fl}^\dagger \rr^\omega & = \frac{2}{\mu_0 \omega^3} \Theta_\text{p} \mathds{A}_\text{HH}^{-1}   \boldsymbol{\chi}^\text{H}   \mathds{A}_\text{HH}^{-1 \dagger},
\end{align}
we obtain 
\begin{align}
  P_{\omega}^\text{np} & = 4 \Theta_\text{p} \mu_0 \omega \epsilon_{ijz} \int \rd^2 x\, \notag \\
& \quad \times \Re \Bigl[ \mathds{G}_{\text{E} k} (\mathbf{r}, \mathbf{r}_\text{p}) \Bigl[ \boldsymbol{\chi}^k + \mathds{F}_k \boldsymbol{\chi}^{\bar{k}} \mathds{F}_k^\dagger \Bigr] \mathds{G}_{\text{H} k}^\dagger (\mathbf{r}, \mathbf{r}_\text{p}) \notag \\
& \quad + \mathds{G}_{\text{E} k} (\mathbf{r}, \mathbf{r}_\text{p}) \Bigl[ \boldsymbol{\chi}^k \mathds{F}_{\bar{k}}^\dagger + \mathds{F}_k \boldsymbol{\chi}^{\bar{k}} \Bigr] \mathds{G}_{\text{H} \bar{k}}^\dagger (\mathbf{r}, \mathbf{r}_\text{p}) \Bigr]_{ij}.
\end{align}
Hence, we can write
\begin{equation}
	P_\omega^\text{np} = - P_\omega^\text{eq}|_{T_\text{s} = T_\text{p}}.
\end{equation}
This is a logical result because in global thermal equilibrium $P_\omega^{\rm s} = P_\omega^{\rm leq} = 0$ holds so that $P_\omega^\text{np} = - P_\omega^\text{eq}$ must be fulfilled to have $P_\omega^{\rm tot} = 0$. In appendix \ref{App:Comparision} we compare the thermal emission $P_\omega^{\rm tot} = P_\omega^\text{np}  +  P_\omega^\text{eq}$ for the purely electric case and $T_\text{s} = T_\text{p}$ with previously obtained results.

\subsection{LEQC}

Similarly, we obtain for the LEQC
\begin{equation}
  P_\omega^\text{leq} = P_{\omega,1}^\text{leq} + P_{\omega,2}^\text{leq} 
\end{equation}
with 
\begin{equation}
  \begin{split}
	  P_{\omega,1}^\text{leq} & = k_0^3 \left( \Theta_\text{b} - \Theta_\text{s} \right) \Im \biggl[ 2 \sqrt{\frac{\varepsilon_0}{\mu_0}} \alpha_\text{HE} R_{c}^\text{pr} \\
                       & \quad + \sum_{j \in \{\perp,z\}} \left( \alpha_{\text{EE},j} R_{\text{E},j}^\text{pr} + \frac{\varepsilon_0}{\mu_0} \alpha_{\text{HH},j} R_{\text{H},j}^\text{pr} \right) \biggr]
  \end{split}
\end{equation}
and the integrals ($k = \text{H,E}$)
\begin{align}
  R_{k,\perp}^\text{pr} & = \int_0^{k_0} \frac{\text{d} k_\perp}{2 \pi} \frac{k_\perp}{k_0 k_z} \Bigl[ \frac{k_z^2}{k_0^2} (1 - |r_k|^2) \left[ 1 - r_k e^{2 \text{i} k_z d} \right] \notag \\
     & \quad + (1 - |r_{\bar{k}}|^2) \left[ 1 + r_{\bar{k}} e^{2 \text{i} k_z d} \right] \Bigr] , \\
  R_{k,z}^\text{pr} & = \int_0^{k_0} \frac{\text{d} k_\perp}{2 \pi} \frac{k_\perp^3}{k_0^3 k_z} (1 - |r_k|^2) \left[ 1 + r_k e^{2 \text{i} k_z d} \right] , \\
  R_{c}^\text{pr} & = \int_0^{k_0} \frac{\text{d} k_\perp}{2 \pi} \frac{k_\perp}{k_0^2} \Bigl[ (1 - |r_\text{H}|^2) r_\text{H} - (1 - |r_\text{E}|^2) r_\text{E} \Bigr] e^{2 \text{i} k_z d}
\end{align}
as well as
\begin{align}
  P_{\omega,2}^\text{leq} & = \frac{k_0^6}{8} \left( \Theta_\text{s} - \Theta_\text{b} \right) \Re \biggl[ \Bigl(\Gamma_{\text{E},\perp} |\alpha_{\text{EE},\perp}|^2 + \frac{\varepsilon_0}{\mu_0} \Gamma_{\text{H},\perp} |\alpha_\text{HE}|^2 \notag \\
    & \quad + 2 \sqrt{\frac{\varepsilon_0}{\mu_0}} \Gamma_{c} \alpha_{\text{EE},\perp} \alpha_\text{HE}^{*} \Bigr) I_{\text{E}, \perp}^\text{pr} + 2 \Gamma_{\text{E},z} |\alpha_{\text{EE},z}|^2 I_{\text{E},z}^\text{pr} \notag \\
   & \quad + \frac{\varepsilon_0^2}{\mu_0^2} \Bigl[ \Bigl( \Gamma_{\text{H},\perp} |\alpha_{\text{HH},\perp}|^2 - 2 \sqrt{\frac{\mu_0}{\varepsilon_0}} \Gamma_{c}^{*} \alpha_{\text{HH},\perp} \alpha_\text{HE}^{*} \notag \\
   & \quad + \frac{\mu_0}{\varepsilon_0} \Gamma_{\text{E},\perp} |\alpha_\text{HE}|^2 \Bigr) I_{\text{H},\perp}^\text{pr} + 2 \Gamma_{\text{E},z} |\alpha_{\text{HH},z}|^2 I_{\text{H},z}^\text{pr} \Bigr] \notag \\
   & \quad + 2 \sqrt{\frac{\varepsilon_0}{\mu_0}} \Bigl[ - \Gamma_{\text{E},\perp} \alpha_{\text{EE},\perp} \alpha_\text{HE}^{*} + \frac{\varepsilon_0}{\mu_0} \Gamma_{\text{H},\perp}^{*} \alpha_{\text{HH},\perp}^{*} \alpha_\text{HE} \notag \\
   & \quad + \sqrt{\frac{\varepsilon_0}{\mu_0}} \Bigl( \Gamma_{c} \alpha_{\text{EE},\perp} \alpha_{\text{HH},\perp}^{*} - \Gamma_{c}^{*} |\alpha_\text{HE}|^2 \Bigr) \Bigr] I_{\text{c}}^\text{pr} \biggr]
\end{align}
with the integrals  ($k = \text{H,E}$)
\begin{align}
   \Gamma_{k,\perp} & = \int_0^{k_0} \frac{\text{d} k_\perp}{2 \pi} \frac{k_\perp}{k_0 k_z} \left(\frac{k_z^2}{k_0^2} \left( 1 - |r_k|^2 \right) + 1 - |r_{\bar{k}}|^2 \right) \notag \\
       & \quad + \int_{k_0}^\infty \frac{\text{d} k_\perp}{\pi} \frac{k_\perp e^{- 2 |k_z| d}}{k_0 |k_z|} \Im\left(\frac{|k_z|^2}{k_0^2} r_k + r_{\bar{k}} \right) , \\
  \Gamma_{k,z} & = \int_0^{k_0} \frac{\text{d} k_\perp}{2 \pi} \frac{k_\perp^3}{k_0^2 k_z} \left( 1 - |r_k|^2 \right) \notag \\
     & \quad + \int_{k_0}^\infty \frac{\text{d} k_\perp}{\pi} \frac{k_\perp^3}{k_0^2 |k_z|} e^{- 2 |k_z| d} \Im(r_k) , \\
   \Gamma_{c} & = \int_0^{k_0} \frac{\text{d} k_\perp}{2 \pi} \frac{k_\perp}{k_0^2} (2 - |r_\text{H}|^2 - |r_\text{E}|^2) \notag \\
      & \quad + \int_{k_0}^\infty \frac{\text{d} k_\perp}{\ri \pi} \frac{k_\perp}{k_0^2} e^{- 2 |k_z| d} \left( \Im(r_\text{H}) - \Im(r_\text{E}) \right) .
\end{align}
One can expect the first part $P^{\rm leq}_{\omega,1}$ to be negative for $T_\text{b} < T_\text{s}$. Since it is proportional to the imaginary parts of the polarizabilities, it can be interpreted as that fraction of the thermal emission of the substrate which is absorbed or shielded by the nanoparticle. We have checked that by adding, for example, the power $P_\omega^{\rm s}$ from Eq.~(\ref{Eq:Psubstrate}) emitted by the substrate for a surface area $A$ which equals the cross section of the nanoparticle, then $P_\omega^{\rm s} + P_{\omega,1}^{\rm leq}$ would be positive for all frequencies. The second part $P^{\rm leq}_{\omega,2}$ which is proportional to the square of the polarizabilities describes, in contrast to $P_{\omega,1}^\text{leq}$, the power of the substrate scattered by the nanoparticle into the far field. The $\Gamma$-integrals are well known from the heat flux received by a small sphere from a semi-infinite substrate when neglecting multiple reflections~\cite{Krueger,Herz2}.

\section{Numerical evaluation}

\begin{figure*}[hbt]
    \center
    \includegraphics[width=0.9\textwidth]{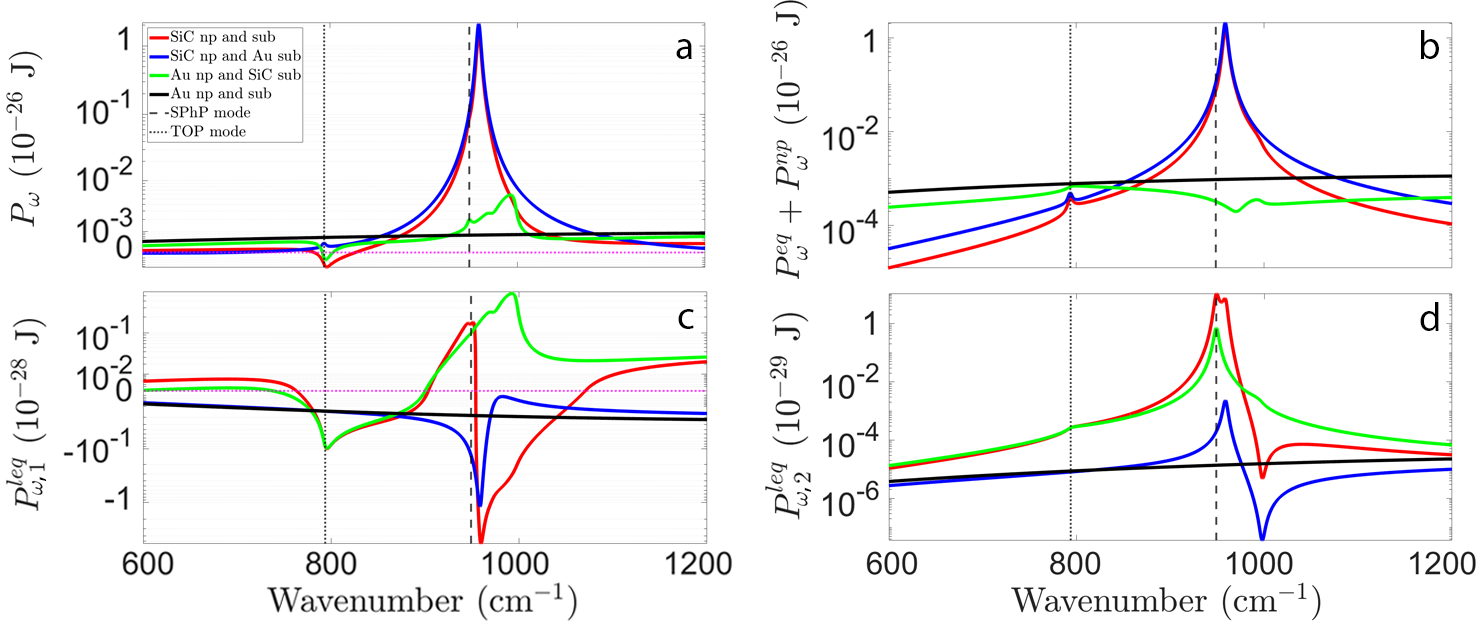}
    \caption{Semi-logarithmic plot of the z-component of spectral power signals for $R = 10$ nm, $d = 40$ nm, $T_\text{p} = 700$ K, $T_\text{s} = 500$ K, and $T_\text{b} = 300$ K for all configurations of gold and SiC as materials of nanoparticle and substrate. The vertical dashed lines indicate the SPhP and TOP wavenumber. Panel (a) shows the spectrum of the total signal, panel (b) the EQC and $P^\text{np}_\omega$, and panel (c,d) the two LEQCs.}
    \label{fig:P_total}
\end{figure*}

Alltogether the total spectral power emitted into the far-field is given by
\begin{equation}
	P_\omega^{\rm tot} = P_\omega^{\rm s} + (P_\omega^{\rm eq} + P_\omega^{\rm np}) + P_{\omega,1}^{\rm leq} + P_{\omega,2}^{\rm leq} 
\end{equation}
As already mentioned $P_\omega^{\rm s}$ does not depend on the presence of the nanoparticle and is therefore not of interest in the following discussion. 
To evaluate these formulas for the emitted power, we choose to compare four different configurations where nanoparticle and substrate can consist of the dielectric material SiC or the metal gold. The permittivity of SiC is modelled by a Lorentz oscillator~\cite{Palik}
\begin{equation}
  \varepsilon_\text{SiC} (\omega)  = \varepsilon_\infty \frac{\omega_l^2 - \omega^2 - \ri \Gamma \omega}{\omega_t^2 - \omega^2 - \ri \Gamma \omega}
\end{equation}
with $\varepsilon_\infty = 6.7$, $\omega_l = 1.827 \times 10^{14} \text{ rad} \text{ s}^{-1}$, $\omega_t = 1.495 \times 10^{14} \text{ rad} \text{ s}^{-1}$, and $\Gamma = 0.9 \times 10^{12} \text{ rad} \text{s}^{-1}$. For gold we employ the Drude model~\cite{Ordal}
\begin{equation}
  \varepsilon_\text{Au} (\omega)  = \varepsilon_\infty - \frac{\omega_p^2}{\omega^2 + \ri \Gamma \omega}
\end{equation}
with $\varepsilon_\infty = 8.344$, $\omega_p = 1.372 \times 10^{16} \text{ rad} \text{ s}^{-1}$, and $\Gamma = 4.059 \times 10^{13} \text{ rad} \text{ s}^{-1}$. Both are necessary for $\chi_k$. For isotropic spherical particles those material constants are given by~\cite{DedkovKyasov07,Dong}
\begin{align}
   \chi_\text{E} & = \frac{9 \ri}{2 k_0^3} \frac{\varepsilon_{\rm p} j_1(y) [x j_1(x)]' - j_1(x) [y j_1(y)]'}{\varepsilon_{\rm p} j_1(y) [x h_1^{(1)}(x)]' - h_1^{(1)}(x) [y j_1(y)]'} , \\
   \chi_\text{H} & = \frac{9 \mu_0 \ri}{2 \varepsilon_0 k_0^3} \frac{j_1(y) [x j_1(x)]' - j_1(x) [y j_1(y)]'}{j_1(y) [x h_1^{(1)}(x)]' - h_1^{(1)}(x) [y j_1(y)]'} 
\end{align}
with $x = k_0 R$ and $y = \sqrt{\varepsilon_\text{p}} x$. $j_1$ and $h_1^{(1)}$ are the spherical Bessel and Hankel functions of the first kind, respectively. The prime notation indicates the derivation with respect to the considered argument. For small radii $R$, meaning $x,y <\!\!\!< 1$, these two expressions can be simplified within the LWA by
\begin{align}
  \chi_\text{E} & = 3 \frac{\varepsilon_\text{p}-1}{\varepsilon_\text{p}+2}, 
  \label{eq:chi_e_approx} \\
  \chi_\text{H} & = \frac{\mu_0}{\varepsilon_0} \frac{(k_0 R)^2}{10} (\varepsilon_\text{p} - 1) .
  \label{eq:chi_m_approx}
\end{align}
Note, that $\Im(\chi_\text{H})$ describes the absorptivity of a small nanoparticle due to the induction of eddy currents as discussed in detail in Refs.~\cite{HuthEtAl2010,Tomchuk}. Therefore, our general approach introducing the magnetic response $\chi_\text{H}$ is by virtue of the correct form of $\chi_\text{H}$ including the contributions of eddy currents which are important for metallic nanoparticles.

\subsection{Spectral signatures}

To ensure that the dipole approximation is applicable, the nanoparticle with $R = 10$ nm is placed above the substrate at distance $d = 40$ nm. All three components have different temperatures: $T_\text{p} = 700$ K, $T_\text{s} = 500$ K, and $T_\text{b} = 300$ K. In the plots we refer to the nanoparticle by ``np'' and to the substrate by ``sub''. Additionally, we highlight the surface mode resonance (SPhP) wavenumber and the transversal optical phonon (TOP) wavenumber of the SiC substrate by vertical dashed lines. To employ semi-logarithmic plots even for negative values, as necessary for $P_{\omega,1}^{\rm leq}$, we adopt the method outlined by Webber \cite{Webber}. 

In Fig.~\ref{fig:P_total}(a) we show the total spectral power without the power emitted by the semi-infinite material $ P_\omega^{\rm s}$ from Eq.~(\ref{Eq:Psubstrate}), i.e.\ we show $P_\omega = P_\omega^{\rm tot} - P_\omega^{\rm s} = P^{\rm np}_\omega + P^{\rm eq}_\omega + P^{\rm leq}_{\omega,1} + P^{\rm leq}_{\omega,2}$. It can be seen that $P_\omega$ is negative for SiC substrates at the TOP wavenumber and, partly, for a SiC nanoparticle above a gold substrate outside the reststrahlen band. For SiC nanoparticles the SPhP mode resonance dominates the spectrum. To see where these effects are coming from, we show the contributions $P_\omega^{\rm eq} + P_\omega^{\rm np}$, $P_{\omega,1}^{\rm leq}$, and $P_{\omega,2}^{\rm leq}$ in Fig.~\ref{fig:P_total}(b)-(d), separately. 

\begin{figure*}[hbt]
    \centering
    \includegraphics[width=0.9\textwidth]{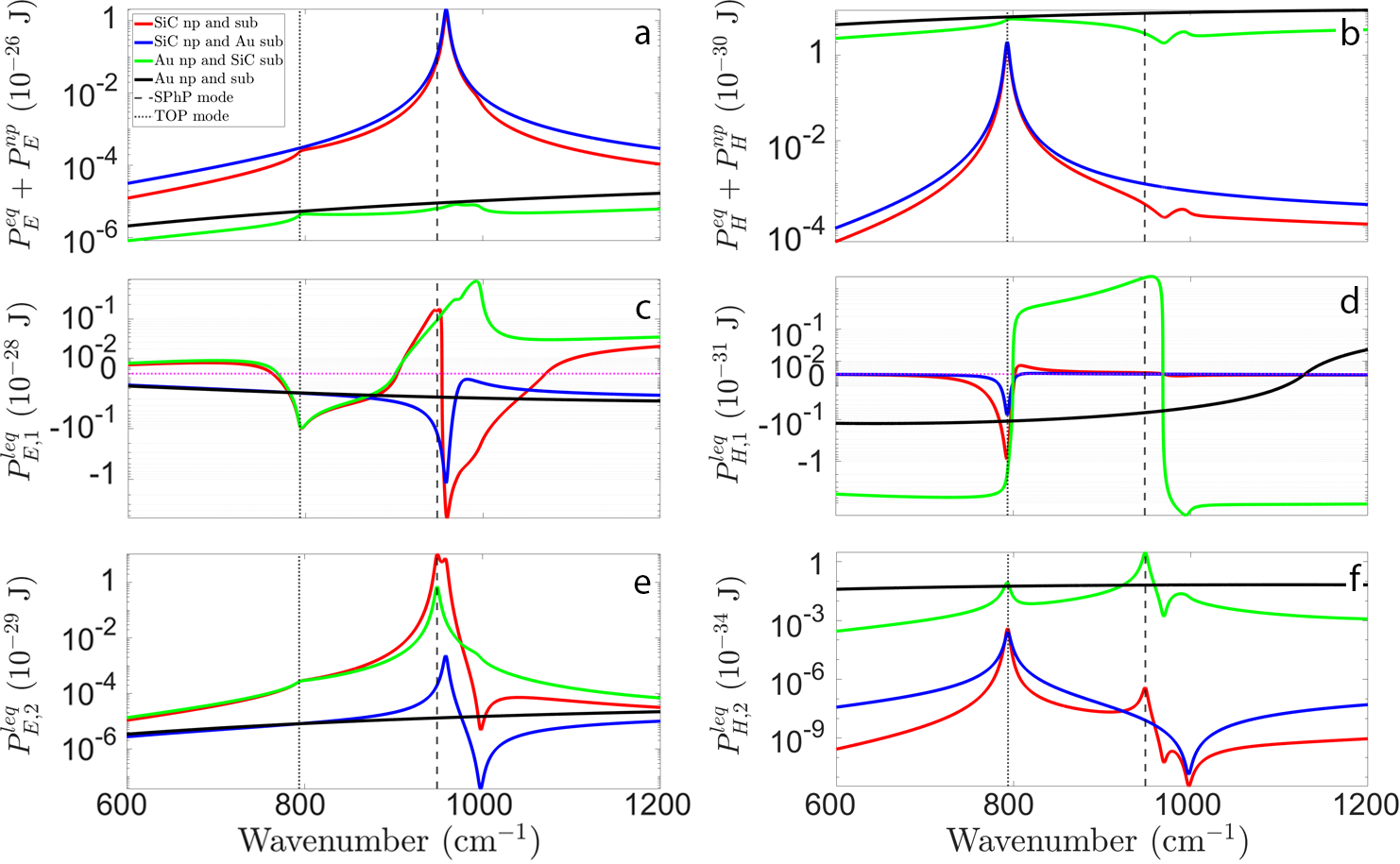}
	\caption{Semi-logarithmic plot of the z-component of the electric (left) and magnetic (right) spectral power. (a,b) show the EQC and $P_\omega^\text{np}$, (c,d) and (e,f) show the two parts of the LEQC. Parameters are the same as in Fig.~\ref{fig:P_total}.}
    \label{fig:P_EH}
\end{figure*}

The quantity $P_\omega^{\rm eq} + P_\omega^{\rm np}$ is plotted in Fig.~\ref{fig:P_total}(b) corresponding to the total spectral power emitted in the case where $T_\text{s} = T_\text{b}$. This power is positive and because of $T_\text{p} \gg T_\text{b}$ it is mainly emitted by the nanoparticle. The spectra are showing resonances for the SiC nanoparticle close to the TOP and SPhP wavenumbers. The pronounced peaks near the SPhP wavenumber are blueshifted due to the strong coupling between the nanoparticle and the substrate. Large shifts of the SPhP resonance frequency -- but towards lower frequencies because of different material compositions -- were already encountered theoretically~\cite{Joulain2,Edalatpour2,Herz} and observed experimentally~\cite{Babuty,Jones}. For the Au nanoparticle above the Au substrate there are, of course, no resonances in the infrared. Nonetheless, even for Au nanoparticles the TOP and SPhP resonances of the SiC substrate have, due to the interaction between the nanoparticle and the substrate, some impact on the emitted power. Since the metal substrate is more reflective than the SiC substrate, the signals emitted by the nanoparticles into the far field are larger for the metal substrates than for SiC substrates, because for Au the part radiated towards the substrate is reflected into the far-field whereas for SiC this radiation is partially absorbed by the substrate. This explains also the dip in the power spectrum at the SPhP wavenumber for the Au nanoparticle above the SiC substrate.

Now, Fig.~\ref{fig:P_total}(c) and (d) show $P_{\omega,1}^{\rm leq}$ and $P_{\omega,2}^{\rm leq}$ which would be the power emitted in the case $T_\text{p} = T_\text{b}$ when neglecting, again, the directly emitted power by the substrate $P^\text{s}_\omega$. As discussed above, $P_{\omega,1}^{\rm leq}$ can be understood as that power of the substrate emission $P_\omega^{\rm s}$ which is absorbed in the nanoparticle due to the multiple interactions between the particle and the substrate. As can be seen in Fig.~\ref{fig:P_total}(c) this contribution can be negative which is as mentioned in Sec.~III.D due to the omisson of the substrate contribution. Therefore, it is not astonishing that $P_{\omega,1}^{\rm leq}$ has a resonance with negative value at the SPhP wavenumber for the SiC nanoparticle where the thermal radiation coming from the substrate is strongly absorbed. For the SiC substrate there is also a resonance with negative value at the TOP wavenumber which simply means that also a part of the power emitted by the surface is reflected by the nanoparticle and then reabsorbed by the substrate. The scattered power $P_{\omega,2}^{\rm leq}$, on the other hand, is always positive and the scattering is particluarly strong around the SPhP wavenumber if the particle or the substrate are made of SiC.

Let us conclude the investigation of the spectra by ascertaining the impact of the electric and magnetic part for each configuration and the different contributions $P_\omega^{\rm eq} + P_\omega^{\rm np}$, $P_{\omega,1}^{\rm leq}$, and $P_{\omega,2}^{\rm leq}$ shown in Fig.~\ref{fig:P_EH}. We do not show the impact of the mixed terms because they are negligibly small compared to the other purely electric or magnetic ones. It is interesting that the resonant thermal emission and scattering near the SPhP mode stem from the electric part due to resonance in $r_\text{E}$ and the peak corresponding to the TOP mode stems from the magnetic part [compare Fig. \ref{fig:P_EH} (a) and (b) and Fig. \ref{fig:P_EH} (e) and (f)]. Because of the higher magnitudes of the electric part, the SPhP mode dominates the spectrum in Fig.~\ref{fig:P_total}a for SiC nanoparticles. Compared to the electric contribution, the magnetic one is by two orders of magnitudes smaller for the SiC nanoparticle and about one order of magnitude larger for the gold nanoparticle. Thus, as could be expected, the magnetic part has a bigger impact on the signal for metallic than for dielectric nanoparticles. This can be in particularly observed in the magnetic emitted power $P_\omega^{\rm eq} + P_\omega^{\rm np}$ in Fig. \ref{fig:P_EH}(b) and in the magnetic scattered power $P_{\omega,2}^{\rm leq}$ shown in Fig.~\ref{fig:P_EH}(f).

\subsection{Temperature and size dependence}

\begin{figure}[t]
    \centering
    \includegraphics[width=0.48\textwidth]{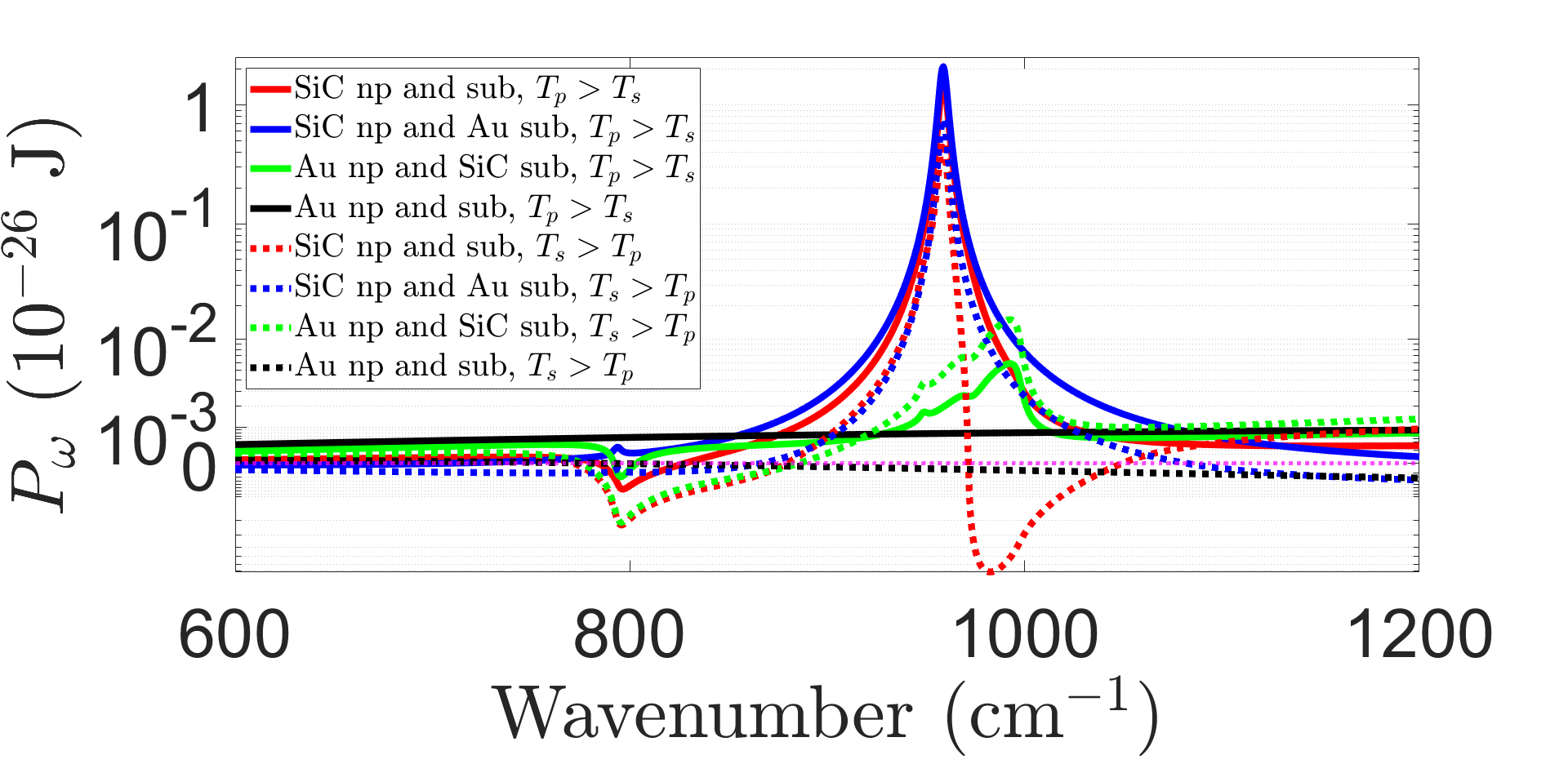}
	\caption{Semi-logarithmic plot of the z-component of spectral power for $R = 10$ nm, $d = 40$ nm, $T_\text{p} = 700$ K, $T_\text{s} = 500$ K, and $T_\text{b} = 300$ K ($T_{\rm p} > T_{\rm s}$, solid lines) for all configurations of gold and SiC as materials of nanoparticle and substrate as in Fig.~\ref{fig:P_total} and with the switched temperatures $T_\text{s} = 700$ K, and $T_\text{p} = 500$ K ($T_{\rm s} > T_{\rm p}$, dashed lines).}
    \label{fig:P_temp}
\end{figure}

Loosely speaking we have the proportionalities
\begin{align}
	P_\omega^{\rm eq} + P_\omega^{\rm np} &\propto (T_{\rm p} - T_{\rm b}) \Im(\alpha), \\
        P_{\omega,1}^{\rm leq} &\propto (T_{\rm b} - T_{\rm s}) \Im(\alpha), \\
        P_{\omega,2}^{\rm leq} &\propto (T_{\rm b} - T_{\rm s}) |\alpha|^2,
\end{align}
where depending on whether the magnetic or electric part dominates one has to consider $\alpha_{\rm EE}$ or $\alpha_{\rm HH}$. The cross terms can typically be neglected. That means that the spectra are highly dependent on the temperature and the radius of the nanoparticle. For instance, by switching the temperatures from $T_p = 700\,{\rm K}$ and $T_{\rm s} = 500\,{\rm K}$ as in Fig.~\ref{fig:P_total} to  $T_p = 500\,{\rm K}$ and $T_{\rm s} = 700\,{\rm K}$ one obtains the results shown in Fig.~\ref{fig:P_temp}. It can be clearly seen that for all material combinations the spectra highly depend on the choice of temperature. It is also clear, that the signs of the different contributions $P_\omega^{\rm eq} + P_\omega^{\rm np}$, $ P_{\omega,1}^{\rm leq}$, and $ P_{\omega,2}^{\rm leq}$ highly depend on the relative value of the different temperatures in the system. In addition, in an experimental setup being able to control $T_{\rm s}$, $T_{\rm b}$, and $T_{\rm p}$ separately it would be possible to measure only $ P_{\omega,1}^{\rm leq}$ and $ P_{\omega,2}^{\rm leq}$ by choosing $T_{\rm p} = T_{\rm b}$ or only $P_\omega^{\rm eq} + P_\omega^{\rm np}$ by choosing $T_{\rm s} = T_{\rm b}$.

\begin{figure}[t]
    \centering
    \includegraphics[width=0.48\textwidth]{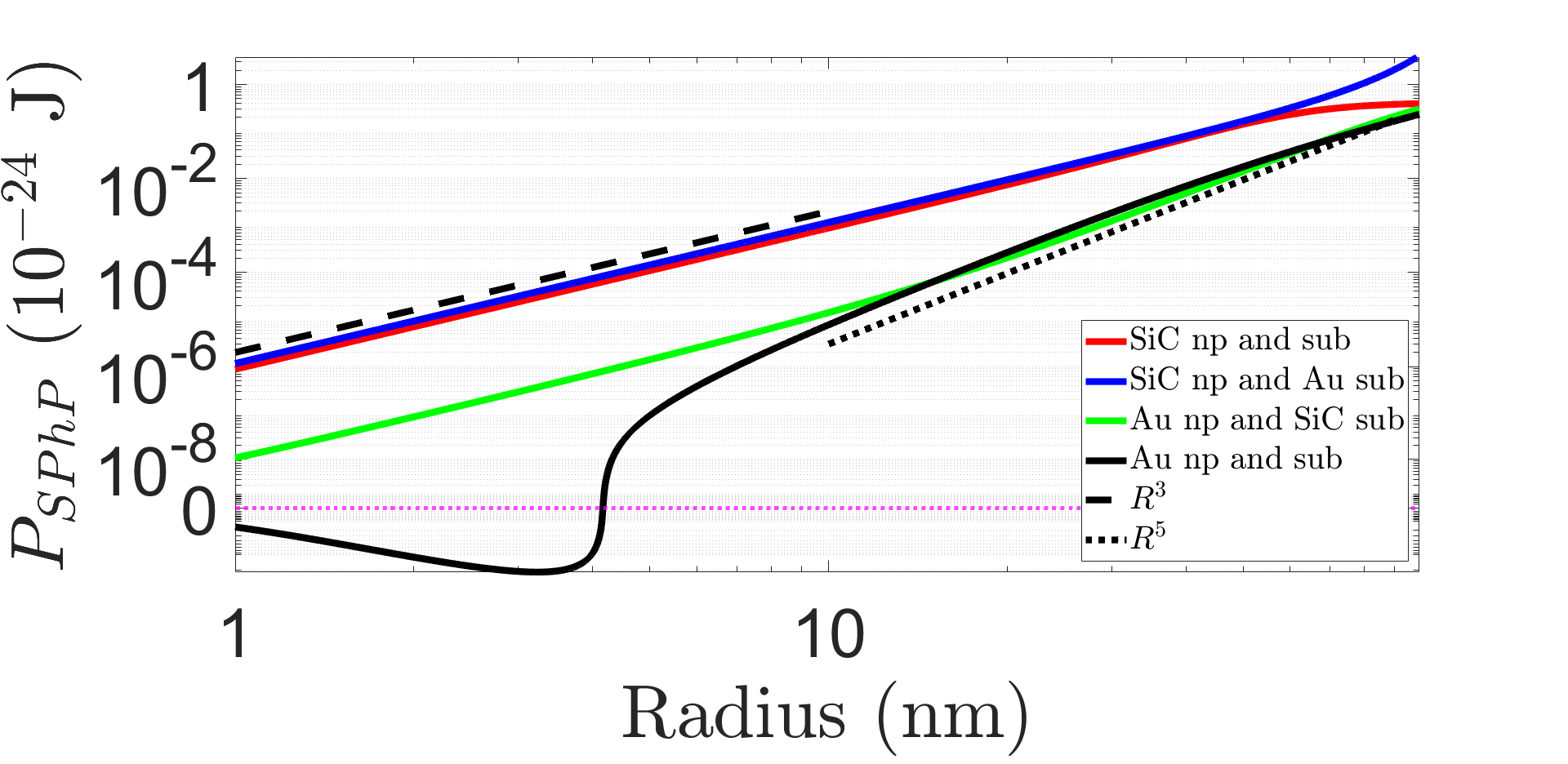}
	\caption{Semi-logarithmic plot of the z-component of spectral power as function of the radius for $d = 100$ nm, $T_\text{p} = 700$ K, $T_\text{s} = 500$ K, and $T_\text{b} = 300$ K for all configurations of gold and SiC as materials of nanoparticle and substrate as in Fig.~\ref{fig:P_total} at the surface mode wavenumber. The dashed line indicates an $R^3$ and the dotted line an $R^5$ power law.}
    \label{fig:P_rad}
\end{figure}

	Since for small radii the polarizabilites fulfill $\alpha_{\rm EE} \propto R^3$ and $\alpha_{\rm HH}  \propto R^5$ it is clear from the above proportionalities that the electric part of $P_\omega^{\rm eq} + P_\omega^{\rm np},  P_{\omega,1}^{\rm leq}$ is $\propto R^3$ and the magnetic part is $\propto R^5$ whereas the electric part of $P_{\omega,2}^{\rm leq}$ is $\propto R^6$ and the magnetic part is $\propto R^{10}$ suggesting that for large enough radii always the scattering term will dominate as long as $T_{\rm s} \neq T_{\rm b}$~\cite{Herz}. In Fig.~\ref{fig:P_rad} we explicitely show the radius dependence for small radii when $d = 100\,{\rm nm}$. It can be seen that for radii smaller than $10\,{\rm nm}$ the emitted power scales like $R^3$ for the dielectric and magnetic nanoparticle whereas for larger radii the magnetic contribution starts to dominate for the metallic nanoparticle leading to a $R^5$ dependence. Of course these dependencies highly depend on the material parameters and the particular dominance of either $P_\omega^{\rm eq} + P_\omega^{\rm np},  P_{\omega,1}^{\rm leq}$ or $  P_{\omega,2}^{\rm leq}$.

\subsection{Angle dependence}

\begin{figure}[t]
    \centering
    \includegraphics[width=0.48\textwidth]{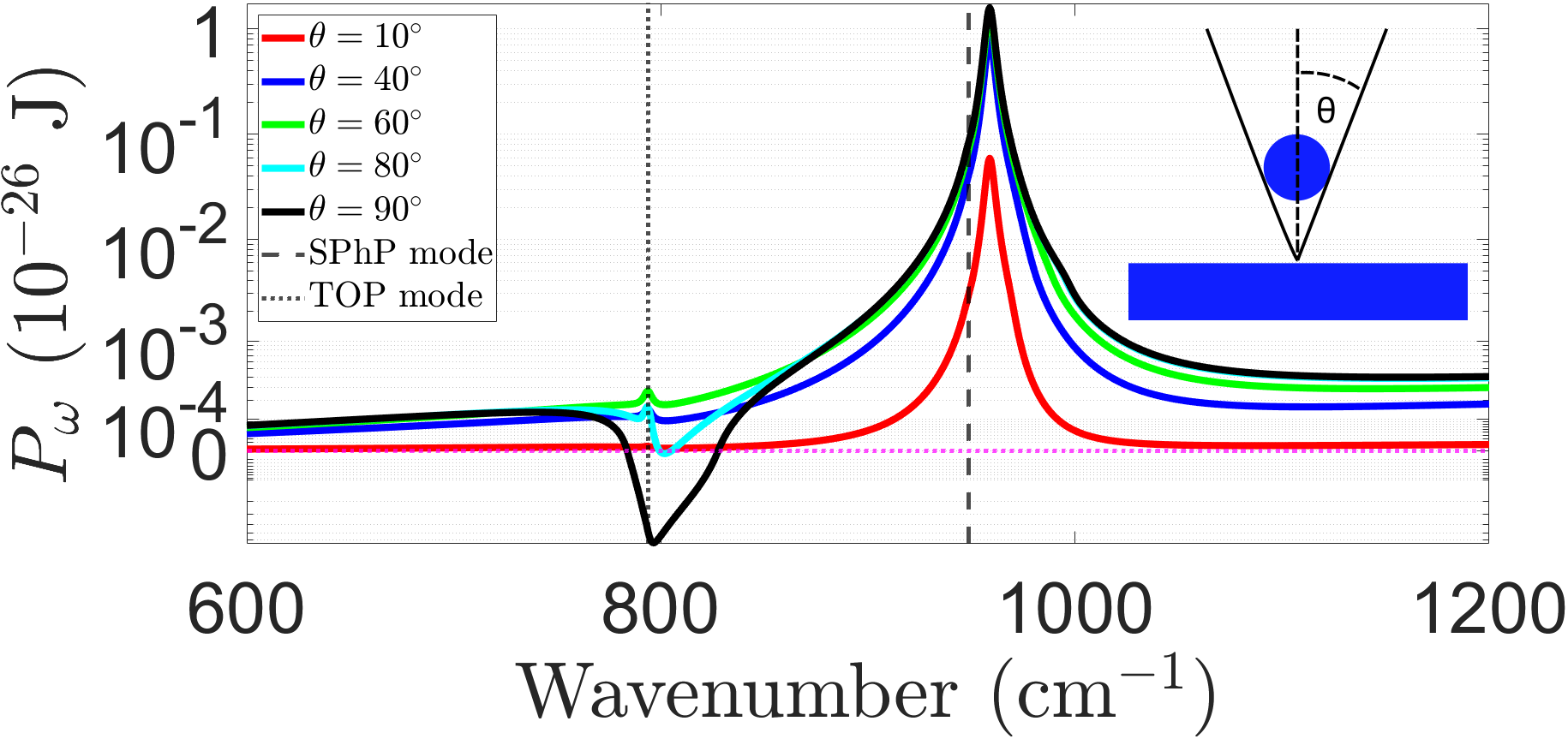}
	\caption{Semi-logarithmic plot of the z-component of the spectral power as function of the opening angle $\theta$ for a SiC nanoparticle above a SiC substrate. Inset: Definition of the angle $\theta$. Parameters are the same as in Fig.~\ref{fig:P_total}}
    \label{fig:P_angle}
\end{figure}

Similar to the approach in Ref.~\cite{Joulain2} our general expressions obtained by integration of the mean Poynting vector over the whole x-y plane can still be used to obtain also the emission into a certain solid angle. That is because the integrals over $\mathbf{k}_\perp$ are taking the emission into all possible directions in the half-space above the nanoparticle into account. To get the angle-dependent expressions, the $k_\perp$ integrals for the quantitities $P_{\omega}^{\rm s}, I^{\rm pr}_{k,\perp},  I^{\rm pr}_{k,z},  I^{\rm pr}_{c}, R^{\rm pr}_{k,\perp},  R^{\rm pr}_{k,z},  R^{\rm pr}_{c}$ must be converted following the scheme
\begin{equation}
\begin{split}
	\int_0^{k_0} \!\! \frac{\rd^2 \mathbf{k}_\perp}{(2 \pi)^2} f(\mathbf{k}_\perp) &= \int_0^{k_0} \!\! \frac{\rd k_\perp}{2 \pi}  \, k_\perp f(k_\perp) \\  &=  \int_0^{\pi/2} \!\! \frac{\rd \theta}{2 \pi} \, \cos(\theta) \sin(\theta) k_0^2 f(k_0 \sin(\theta)) 
\end{split}
	\label{Eq:openingAngle}
\end{equation}
where $f$ denotes the integrand of the different quantities and $\theta$ is the angle between $\mathbf{k}_\perp$ and the surface normal as indicated in Fig. \ref{fig:P_angle}. Note, that in our calculations we have already used the rotational symmetry to simplify the two-dimensional integrals to a one-dimensional integral. Therefore, when it is necessary to consider only the thermal emission into a certain solid angle range $\Delta \theta$ and $\Delta \phi$ it is inevitable to make the replacements
\begin{equation}
  \begin{split}
	  \int_0^{k_0} \! & \frac{\rd k_\perp}{2 \pi} k_\perp f(k_\perp) \\ 
	      &\rightarrow \frac{\Delta \phi}{2 \pi} \int_{\Delta \theta} \frac{\rd \theta }{2 \pi} \, \cos(\theta) \sin(\theta) k_0^2 f(k_0 \sin(\theta))
  \end{split}
\end{equation}
in the terms $P_{\omega}^{\rm s}, I^{\rm pr}_{k,\perp},  I^{\rm pr}_{k,z},  I^{\rm pr}_{c}, R^{\rm pr}_{k,\perp},  R^{\rm pr}_{k,z},  R^{\rm pr}_{c}$.

	To illustrate the angle dependence of the spectra we show in Fig.~\ref{fig:P_angle} the dependence of the emitted power spectrum on the opening angle $\theta$ as sketched in the inset for a SiC nanoparticle above a SiC substrate using Eq.~(\ref{Eq:openingAngle}). It is obvious that the spectrum is highly depending on the opening angle. It is interesting that for small angles a resonance appears close to TOP wavenumber which is not present for large angles, whereas for an angle $\theta = 60^\circ$ the spectrum around the surface mode wavenumber corresponds already quite well to the full spectrum. Hence, for a modeling of experiments the inclusion of the correct opening angle over which the scattered radiation is collected is very important.

\subsection{Distance dependence}

\begin{figure*}[t]
    \centering
    \includegraphics[width=0.9\textwidth]{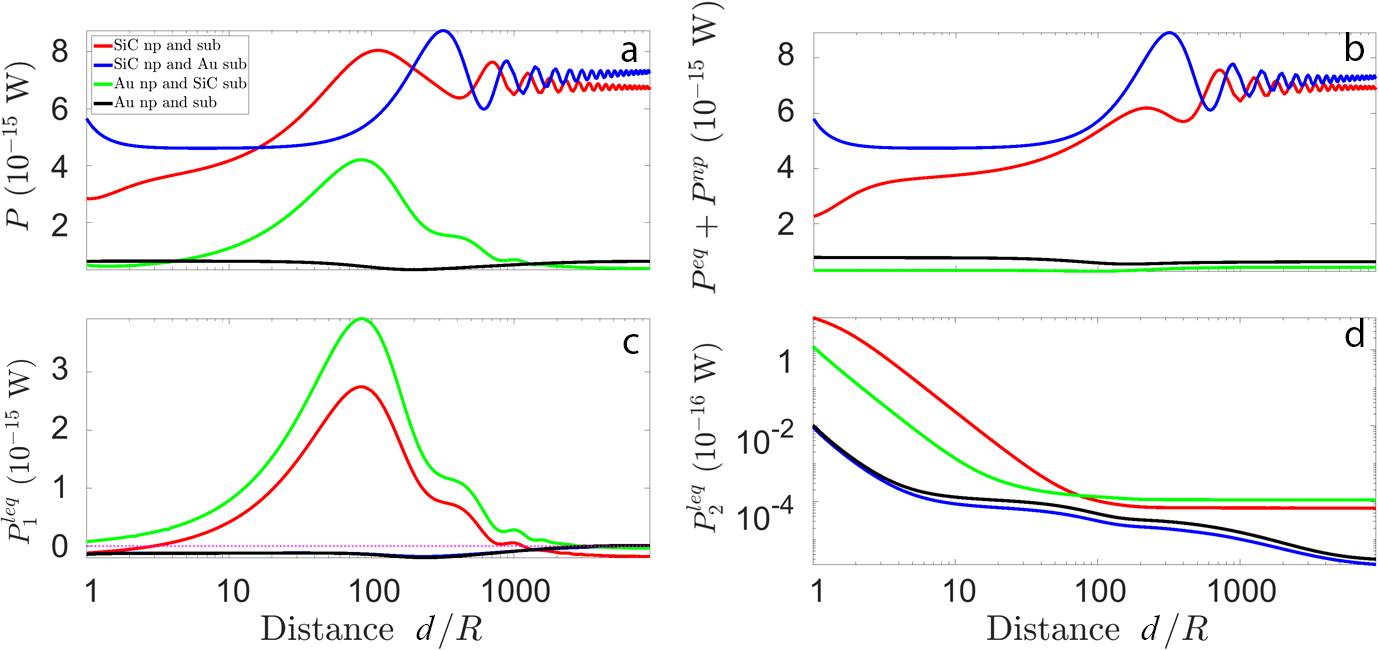}
	\caption{ Logarithmic plot of the total power $P$ without the direct surface emitted power from Eq.~(\ref{Eq.TotalwoS}) as function of distance together with the different contributions $P^{\rm eq} + P^{\rm np}$, $P_1^{\rm leq}$, and $P_2^{\rm leq}$. All other parameters are the same as in Fig.~\ref{fig:P_total}.}
    \label{fig:P_d}
\end{figure*}

Finally, we want to discuss the distance dependence of the total power without the power emitted by the substrate
\begin{equation}
  P = \int_0^\infty \frac{\rd \omega}{2 \pi} (P_\omega^{\rm tot} - P_\omega^{\rm s}) .
\label{Eq.TotalwoS}
\end{equation}
This quantity is plotted over the distance normalized to the particle radius $R$ in Fig.~\ref{fig:P_d} for the four possible material configurations. In Fig.~\ref{fig:P_d} also the different contributions are shown separately. In principle, three different regimes are visible. The extreme near-field ($d \sim R$) where the coupling between the particle and the evanescent waves is dominating. In this regime the scattering contribution $P^{\rm leq}_2$ gives a strong increasement of the heat flux when approaching the medium. On the other hand, such an enhancment cannot be seen in the total heat flux with our set of parameters, since in this case the contribution of $P^{\rm eq} + P^{\rm np}$ dominates for the SiC nanoparticle and $P^{\rm leq}_1$ dominates for the gold nanoparticle. However, by increasing the nanoparticle radius or the substrate temperature $P^{\rm leq}_2$ will eventually dominate the total power emission. The transition region towards the far-field where due to reflections there will be an interference pattern is starting at distances $d \approx 300 R$ which correspond to the thermal wavelength for $T_\text{p} = 700$ K and $R = 10$ nm. The oscillations are exponentially damped so that a constant value is reached in the far-field. Finally, there is an intermediate regime for $2R < d < 300 R$. Due to the high reflectivity, the emitted power (without the direct thermal emission of the substrate) for the Au substrates is larger than for the SiC substrate in the far-field. It is interesting that the oscillations most pronounced for SiC nanoparticles. In the extreme near-field ($d \sim R$) the power emission for the SiC nanoparticle is the largest for $R = 10\,{\rm nm}$. For other choices of parameters and in particular the radius the power emitted by the Au nanoparticle can be larger than that of the SiC nanoparticle for certain configurations as can be seen in Fig.~\ref{fig:P_rad} for radii close to 100 nm. It should, of course, be kept in mind that in the extreme near-field regime also higher multipole orders need to be included~\cite{Naraynaswamy2008,Otey,Becerril} and phonon tunneling might also play a role~\cite{Pendry,GangChen}.

\begin{figure}[t]
    \centering
    \includegraphics[width=0.45\textwidth]{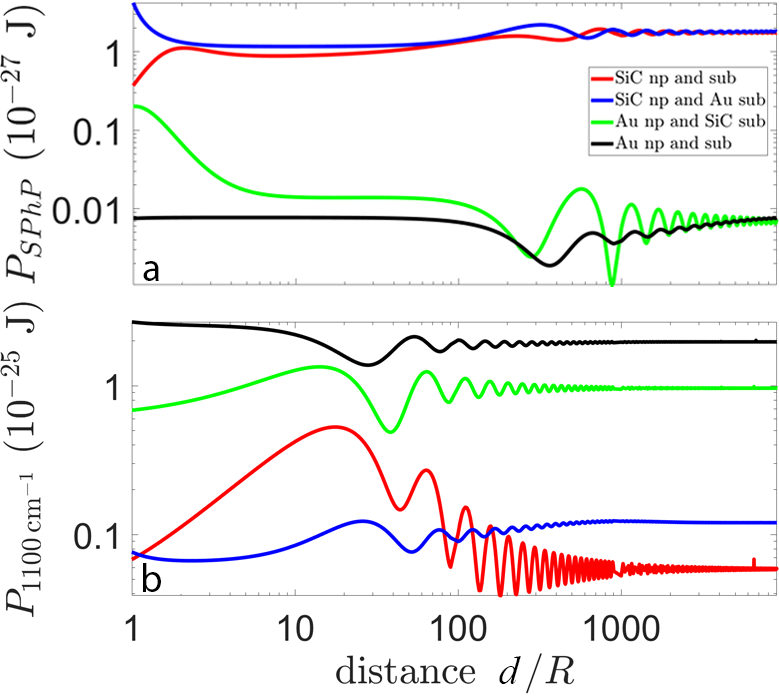}
	\caption{ Logarithmic plot of the spectral power $P_\omega$ as function of distance at the surface mode frequency and the wavenumber $1100\,{\rm cm}^{-1}$. All other parameters are the same as in Fig.~\ref{fig:P_total}.}
    \label{fig:P_d_single}
\end{figure}

	Since scanning thermal microscopes like the TRSTM, TINS, and SNoiM are capable of measuring either spectra or the distance dependence in a narrow frequency band around a single frequency we also show in Fig.~\ref{fig:P_d_single} some results for the distance dependence of the emitted spectral power at the surface mode frequency $\omega_{\rm SPhP}$ and a frequency slightly larger than the longitudinal optical phonon frequency $\omega_l$ of SiC. First of all, it can be seen that the oscillations in the far-field regime are present in all material combinations for single frequencies and therefore are in the integrated power $P$ shown in Fig.~\ref{fig:P_d} albeit smeared out for gold nanoparticles. Furthermore, in the near-field regime an increased or decreased power emission can be seen for the SiC nanoparticle. This effect is particularly strong at the surface mode frequency and survives after integration in the total emitted power for the gold substrate in Fig.~\ref{fig:P_d}. For the Au nanoparticle there can also be an increased power emission when approaching the material depending on the substrate and frequency. Furthermore far away from the surface mode frequency the power emission of the Au nanoparticle can be larger than that of the SiC nanoparticle which, of course, is particularly strong close to the surface mode frequency.    

\section{Conclusion}

We have presented a generalized dipole model within the framework of fluctuational electrodynamics and with a rigurous treatment of the divergence in the Green's functions. Our model is taking eddy currents in the nanoparticle into account as well as the possibility to have three different temperatures for the nanoparticle, substrate, and background. We derived an expression for the spectral power emitted into far-field and discussed numerical emission spectra and the integrated power for all four combinations of the materials SiC and gold for the nanoparticles and the substrate. We have shown that the impact of the three different temperatures can alter the results tremendously. For example by considering the specific cases $T_\text{s} = T_\text{b}$ and $T_\text{p} = T_\text{b}$ either direct thermal emission of the nanoparticle or scattering dominates. The relative values of the three temperatures have a strong impact on the sign of the different contributions. We have discussed the impact of the magnetic field contribution due to the eddy currents in metals. Finally, we discussed the distance dependence of the total power emission without the direct thermal emission of the substrate. A strong near-field increasement can always be seen when the scattering term dominates, whereas in the regime where the direct thermal emission of the nanoparticle dominates there is not necessarily a near-field increasement in thermal emission. Which of both terms dominates the power emission highly depends on the parameters, but one can see from the expressions that for large enough radii scattering always dominates.

In principle, our dipol model can be applied to thermal imaging experiments like TRSTM, TINS, or SNoiM but several further extensions are necessary in upcoming works. To treat, e.g.\ anisotropic or non-reciprocal materials, the evaluation and usage of the Green's functions has to be adapted as well as the definition of the expressions for $\chi_k$ in the Lippmann-Schwinger equation in Eq.~\eqref{eq:lse}. To have a better model for thermal imaging experiments, one could include surface roughness of the surface by perturbation theory \cite{Biehs2010,Chen} or a specific surface profile by combining our model with a numerical method which evaluates the Green's function. Additionally, the exact near-field behavior for $d < 3 R$ requires taking different multipole orders into account instead of relying on the dipole approximation alone. A simple way to improve the modeling of the tip shape used in the near-field imaging experiments could be to extend our work to the case of many dipoles in a discrete dipole approximation \cite{PBAmanybody,Edalatpour2,EkerothEtAl2017}. 

\acknowledgments

S.-A.\ B. acknowledges support from Heisenberg Programme of the Deutsche Forschungsgemeinschaft (DFG, German Research Foundation) under the project No. 404073166.

\appendix
\section{Green's functions}
\label{App:GreenDyads}

The Green's functions for $z > z'$ above a planar, isotropic, and semi-infinite substrate at $z < 0$ are within the Weyl representation~\cite{Hecht,Sipe}
\begin{equation}
\mathds{G}_\text{FF'} (\mathbf{r}, \mathbf{r}') = \int \frac{\text{d}^2 k_\perp}{(2 \pi)^2} e^{\ri \mathbf{k}_\perp (\mathbf{x} - \mathbf{x}')} \mathds{G}_\text{FF'} (k_\perp, z, z')
\end{equation}
with $\mathbf{x} = (x,y)^t$, $\mathbf{x}' = (x',y')^t$, $\mathbf{k}_\perp = (k_x, k_y)^t$, and $\text{F}, \text{F'} = \text{H}, \text{E}$. For example, the electric Green's function for electric sources including the vacuum and scattering contribution is given by~\cite{Hecht,Sipe}
\begin{equation}
  \begin{split}
	  \mathds{G}_\text{EE} (k_\perp, z, z') &= \frac{\ri \re^{\text{i} k_z (z-z')}}{2 k_z} \sum_{k = \text{E,H}} \Bigl( \mathbf{a}_k^{+} (k_0) \otimes \mathbf{a}_k^{+} (k_0) \\ &\qquad + e^{2 \text{i} k_z z'} r_k \mathbf{a}_k^{+} (k_0) \otimes \mathbf{a}_k^{-} (k_0) \Bigr)
  \end{split}
\end{equation}
introducing the polarization vectors~\cite{Hecht,Sipe}
\begin{align}
  \mathbf{a}_\text{H}^{\pm} (k_\mu) & = \frac{1}{k_\perp} \left( k_y, - k_x, 0 \right)^t ,\\
  \mathbf{a}_\text{E}^{\pm} (k_\mu) & = \frac{1}{k_\mu k_\perp} \left( \mp k_x k_{\mu,z}, \mp k_y k_{\mu,z}, k_{\mu,z}^2 \right)^t 
\end{align}
with $k_\mu^2 = \epsilon_\mu \omega^2/c^2$ and $k_{\mu,z}^2 = k_0^2 \epsilon_\mu - \kappa^2$. Here, $r_\text{H/E}$ are the Fresnel amplitude reflection coefficients 
\begin{equation}
  r_\text{H} = \frac{k_z - k_{\text{s},z}}{k_z - k_{\text{s},z}} ,\quad
  r_\text{E} = \frac{\varepsilon_\text{s} k_z - k_{\text{s},z}}{\varepsilon_\text{s} k_z - k_{\text{s},z}} 
\end{equation}
with $k_{\text{s},z}  = \sqrt{\varepsilon_\text{s} k_0^2 - k_\perp^2}$. $\varepsilon_\text{s}$ denotes the substrate permittivity.
 To obtain $\mathds{G}_\text{HH}$ one can simply interchange $r_\text{E} \leftrightarrow r_\text{H}$ and multiply by $\varepsilon/\mu_0$. To obtain $\mathds{G}_\text{HE}$ replace the left-handed vector $\mathbf{a}_\text{E}^{+}$ by $\mathbf{a}_\text{H}^{+}/\omega \mu_0$ and $\mathbf{a}_\text{H}^{+}$ by $-\mathbf{a}_\text{E}^{+}/\omega \mu_0$. One obtains $\mathds{G}_\text{EH}$ from $\mathds{G}_\text{HE}$ by interchanging $r_\text{E} \leftrightarrow r_\text{H}$ and switching the sign.

The electric Green's function for the transmitted part of the fields with electric sources inside the semi-infinite medium ($z' < 0$) is given by the transmitted or scattered part, only, for $z > 0$ and reads
\begin{equation}
\begin{split}
	\mathds{G}_\text{EE}^\text{s} (k_\perp, z, z') & = \frac{\ri \re^{\text{i} (k_z z - k_{\text{s},z} z')}}{2 k_{\text{s},z}} \sum_{k = \text{E,H}} t_k \mathbf{a}_k^{+} (k_0) \otimes \mathbf{a}_k^{+} (k_\text{s})
\end{split}
\end{equation}
introducing the amplitude transmission coefficients
\begin{equation}
  t_\text{H} = \frac{2 k_{\text{s},z}}{k_{\text{s},z} + k_z}, \quad
  t_\text{E} = \frac{k_\text{s}}{k_0} \frac{2 k_{\text{s},z}}{k_{\text{s},z} + k_z} .
\end{equation}
To obtain the other Green's functions, one can perform the same replacements as above.

\section{Volume average of the Green's functions}
\label{App:VolumeAverage}

The volume averaged functions can be separated into a vacuum contribution and a scattered one due to the reflecting substrate. The vacuum contribution for the fully electric Green's function is~\cite{Albaladejo, Fikioris, Lakhtakia, Yaghjian}
\begin{align}
  V \langle \mathds{G}_{\text{EE,vac}} (\mathbf{r}) \rangle & = \frac{1}{k_0^2} \left[ \frac{2}{3} \left( 1 - \ri k_0 R \right) e^{\ri k_0 R} - 1 \right] \mathds{1} .
\label{eq:G_EE_vac}
\end{align}
To obtain the magnetic counterpart, multiply Eq. \eqref{eq:G_EE_vac} with $\varepsilon_0/\mu_0$. The vacuum part of the function $\mathds{G}_\text{EH/HE}$ vanishes. For the considered media the scattered part of the Green's function is well known~\cite{Sipe}. See also appendix~\ref{App:GreenDyads}. Its volume average contains no singularities rendering it possible to approximate the integral as product of the nanoparticle volume and the integrand evaluated at $\mathbf{r}' = \mathbf{r} = d \mathbf{e}_z$ within the long wavelength approximation (LWA). Thus, we obtain 
\begin{align}
  \langle \mathds{G}_\text{EE/HH,sc} \rangle & = \ri \int_0^{\infty} \frac{\text{d} k_\perp}{8 \pi} \frac{k_\perp e^{2 \ri k_z d}}{k_z} \biggl[ \biggl( r_\text{H/E} - r_\text{E/H} \frac{k_z^2}{k_0^2} \biggr) \notag \\
& \quad \times \mathbf{e}_\perp \otimes \mathbf{e}_\perp + 2 \frac{k_\perp^2}{k_0^2} r_\text{E/H} \mathbf{e}_z \otimes \mathbf{e}_z \biggr], \\
  \langle \mathds{G}_\text{EH/HE,sc} \rangle & = - \frac{\ri}{k_0} \sqrt{\frac{\varepsilon_0}{\mu_0}} \int_0^{\infty} \frac{\text{d} k_\perp}{8 \pi} k_\perp e^{2 \ri k_z d} \left( r_\text{H} - r_\text{E} \right) \mathds{X} .
\end{align}

\section{Comparision with previous results for electrical case and $T_\text{s} = T_\text{b}$}
\label{App:Comparision}

If we would restrict ourselves to the purely electric contribution, we have to set $\chi_\text{H}=0$. Then, the polarizability matrices $\boldsymbol{\alpha}_\text{HH/HE}$ contain no non-zero entries anymore and the matrices $\boldsymbol{\alpha}_\text{EE}$ and $\boldsymbol{\chi}^\text{E}$ simplify to
\begin{align}
  \boldsymbol{\alpha}_\text{EE} & = \sum_{j \in \{\perp,z\} } \frac{V \chi_\text{E} \mathbf{e}_j \otimes \mathbf{e}_j}{1 - k_0^2 V \chi_\text{E} \langle G_{\text{EE},j} (\mathbf{r}_\text{p}) \rangle} , \\
  \chi^\text{E}_{\perp/z} & = \frac{k_0^2 V \Im (\chi_\text{E})}{\Big|1 - k_0^2 V \chi_\text{E} \langle G_{\text{EE},\perp/z} (\mathbf{r}_\text{p})  \rangle \Big|^2} .
  \end{align}
Hence, for $T_\text{s} = T_\text{b}$ we would end up with
\begin{align}
P_z^{\rm tot} & = \sum_{j \in \{\perp,z\} } \frac{k_0^3 V \left( \Theta_\text{p} - \Theta_\text{b} \right) \Im (\chi_\text{E}) I_{\text{E},j}^\text{pr}}{\big|1 - k_0^2 V \chi_\text{E} \langle G_{\text{EE},j} (\mathbf{r}_\text{p}) \rangle \big|^2} .
\end{align}
This result resembles the one in Eq.~(62) in Ref.~\cite{Herz} but with $V \chi_\text{e}$ instead of the ``naked'' electric polarizability used there and a different vacuum contribution of the Green's function. Within the range of validity of the dipole model, our new result and that in Ref.~\cite{Herz} give the same numerical values if we neglect the vacuum correction term $\mathds{G}_\text{vac}$.

\section{Substrate as perfect metal}
\label{App:PM}

If the substrate would be a perfect metal ($r_\text{E/H} = \pm 1$), the nanoparticle directly emits
\begin{align}
P_\omega^\text{np} + P_\omega^\text{eq} & = \frac{2 k_0}{3 \gamma^3 \pi} (\Theta_\text{p} - \Theta_\text{b}) \notag \\ 
&\quad \times\Bigl[ \left(\chi_{\text{E},\perp} + |F_\text{E}|^2 \chi_{\text{H},\perp} \right) I_\perp^{-} + \chi_{\text{E},z} I_z^{-} \notag \\
& \quad + \frac{\varepsilon_0}{\mu_0} \left[ \left(\chi_{\text{H},\perp} + |F_\text{H}|^2 \chi_{\text{E},\perp} \right) I_\perp^{+} + \chi_{\text{H},z} I_z^{+} \right] \notag \\
& \quad + 6 \gamma \sqrt{\frac{\varepsilon_0}{\mu_0}} \bigl[ \chi_{\text{H},\perp} \Im(F_\text{E}) + \chi_{\text{E},\perp} \Im(F_\text{H}) \bigr] I \Bigr]
\end{align}
with
\begin{align}
I_\perp^{\mp} & = 2 \gamma^3 \mp 3 \left[\gamma \cos(\gamma) + \left( \gamma^2 - 1 \right) \sin(\gamma) \right] , \\
I_z^{\mp} & = \gamma^3 \mp 3 \left[\gamma \cos(\gamma) - \sin(\gamma) \right] , \\
I & = \sin(\gamma) - \gamma \cos(\gamma) .
\end{align}
Here, we used $\gamma = 2 k_0 d$. In this case the LEQC and $P_\omega^\text{s}$ vanish.

\section{Substrate as black body}
\label{App:BB}

By assuming the substrate to be a black body, we set $r_\text{E/H} = 0$. Then, we get for the directly emitted power of the nanoparticle
\begin{align}
	P_\omega^\text{np} +  P_\omega^\text{eq} & = \frac{k_0}{3 \pi} (\Theta_\text{p} - \Theta_\text{b} ) \left[ 2 \chi_\perp^\text{E} + \chi_z^\text{E} + \frac{\varepsilon_0}{\mu_0} \left( 2 \chi_\perp^\text{H} + \chi_z^\text{H} \right) \right] .
\end{align}
The LEQCs are
\begin{align}
P_{\omega,1}^\text{leq} & = \frac{k_0^3}{3 \pi} \left( \Theta_\text{b} - \Theta_\text{s} \right) \Im \Bigl[ 2 \alpha_{\text{EE},\perp} + \alpha_{\text{EE},z} \notag \\
& \quad + \frac{\varepsilon_0}{\mu_0} \left( 2 \alpha_{\text{HH},\perp} + \alpha_{\text{HH},z} \right) \Bigr] 
\end{align}
and
\begin{align}
P_{\omega,2}^\text{leq} & = \frac{k_0^6}{16 \pi^2} \left( \Theta_\text{s} - \Theta_\text{b} \right) \biggl[ \frac{\varepsilon_0}{\mu_0} \Re \left( \alpha_{\text{EE},\perp} \alpha_{\text{HH},\perp}^{*} \right) \notag \\
& \quad + \frac{8}{9} |\alpha_{\text{EE},\perp}|^2 + \frac{4}{9} |\alpha_{\text{EE},z}|^2 \notag \\
& \quad + \frac{\varepsilon_0^2}{\mu_0^2} \left( \frac{8}{9} |\alpha_{\text{HH},\perp}|^2 + \frac{4}{9} |\alpha_{\text{HH},z}|^2 \right) \biggr] .
\end{align}
In this case $P_\omega^\text{s}$ is simply the well-known power exchanged between two blackbodies at temperatures $T_\text{s}$ and $T_\text{b}$ through an area $A$.

\end{document}